\documentclass[preprint,
showpacs,preprintnumbers,amsmath,amssymb]{revtex4}

\usepackage{graphicx}
\usepackage{dcolumn}
\usepackage{bm,morefloats}
\usepackage{color}

\usepackage{amssymb,bm}

\def\Eqref#1{Eq.~(\ref{#1})}

\def\Eq#1{\begin{equation} #1 \end{equation}}
\def\Eqr#1{\begin{eqnarray} #1 \end{eqnarray}}
\def\Eqrsubl#1#2{\begin{subequations}\label{#1}\Eqr{#2}\end{subequations}}

\newcommand{\nn}{\nonumber}
\newcommand{\pd}{\partial}

\def\Xsp{{\rm X}}

\def\Ysp{{\rm Y}}
\def\Zsp{{\rm Z}}
\def\X5sp{{\rm X}_5}
\def\Y3sp{{\rm Y}_3}
\def\Z3sp{{\rm Z}_3}

\def\Msp{{\rm M}}
\def\Nsp{{\rm N}}
\def\lap{{\triangle}}
\def\e{{\rm e}}

\newcommand{\vect}[1]{\!\!\!\mbox{ \boldmath $#1$}}

\newcommand\bea{\begin{eqnarray}}
\newcommand\eea{\end{eqnarray}}

\begin{document}


\title{
Cosmology in $p$-brane systems
}

\author{Masato Minamitsuji}%
\affiliation{
Department of Physics, 
Graduate School of Science and Technology,
Kwansei Gakuin University, Sanda 669-1337, Japan.
}%

\author{Kunihito Uzawa}
\affiliation{%
Department of Physics, Kinki University,
Higashi-Osaka, Osaka 577-8502, Japan
}%

\date{\today}

\begin{abstract}
We present time-dependent solutions in the higher-dimensional 
gravity which are related to supergravity
in the particular cases. 
Here, we consider $p$-branes with a cosmological constant
and the intersections of 
two and more branes.
The dynamical description of $p$-branes can be naturally  
obtained as the extension of static solutions.
In the presence of a cosmological constant,
we find accelerating solutions 
if the dilaton is not dynamical.
In the case of intersecting branes,
the field equations normally indicate that time-dependent solutions 
in supergravity can be found if only one harmonic function in the metric 
depends on time.
However, if the special relation between dilaton couplings
to antisymmetric tensor field strengths
is satisfied, one can find a new class of solutions
where all harmonic functions 
depend on time. 
We then apply our new solutions 
to study cosmology, 
with and without performing compactifications.
\end{abstract}

\pacs{11.25.-w, 11.27.+d, 98.80.Cq}
\maketitle


\section{Introduction}
 \label{sec:introduction}

The dynamical brane systems in supergravity 
(and more general intersecting brane systems) have attracted
growing interests in recent years since they can be used to construct 
the cosmological model under the compactifications 
in string theory. The simplest dynamical solution 
to Einstein equations in supergravity is a $p$-brane 
in an asymptotically time-dependent background, obtained 
in a system composed of gravity, 
a scalar field and an antisymmetric form field strength. 
In such a solution, 
a naked singularity is formed at the places where
the warp factor vanishes. Such
a solution can be naturally constructed as 
an extension of a static $p$-brane solution. 
This construction has a natural interpretation in terms of D-branes and
has served as an important example in string theory. 
 In the absence of the time dependence, 
a brane system is supersymmetric. 
Solutions can also be constructed 
by lifting 
the Maki-Shiraishi solutions \cite{Maki:1992tq} to higher dimensions. 
These models have interesting 
effects that can spoil 
asymptotic flatness and supersymmetry even if 
they hold in static solutions; 
much attention has been paid on determining 
conditions to obtain a supersymmetric solution 
(for example, see \cite{Maeda:2009tq, Maeda:2010yk}). 
A close cousin of the above solution is a $p$-brane with a
cosmological constant. 
For a single 2-form field strength, 
this is an asymptotically Milne universe. 
The examples relevant for us are the multicentered
Kastor-Traschen solutions \cite{Kastor:1992nn}. 
Some of these developments have been motivated by de Sitter 
compactifications in the four-dimensional effective theory.
Borrowing these results, we acquire a few novel insights 
about the physics. 

In a $p$-brane model, the dynamics can be characterized by
 the warp factor which is given in terms of 
the linear combination of the linear functions of time
and the harmonic function 
in the space transverse to the brane. This function contains 
information about the dynamics of the underlying model, but this has not
been fully exploited yet. 
See \cite{Kodama:2005fz, Binetruy:2007tu, 
Maeda:2009zi, Minamitsuji:2010kb}
 for an example of determination of such a function. 
Since a warp factor arises from a field strength, 
the dynamics of a system composed of $n$ branes can be 
characterized by $n$ warp factors 
arising from $n$ field strengths \cite{Ohta:1997wp, Ohta:1997wd, 
Ohta:2003rr, Miao:2004bn, Chen:2005uw, Ohta:2006sw, Aref'eva:1997nz, 
Maeda:2009ds, Maeda:2010ja}.
In such a system, some of branes can
naturally intersect. However, for M-branes and D-branes, 
among these warp factors only one function can depend on time. 
These harmonic functions of D-brane model are 
related to the string coupling constant in string theory.
They have been studied from many points of view; 
for recent discussion, see \cite{Gibbons:2009dr, Nozawa:2010zg}.

The purpose of the present paper is to make this result more 
transparent and to generalize it. We will consider solutions 
with more general couplings 
of dilaton to the field strengths.
In the classical solution of a $p$-brane in 
a $D$-dimensional theory, 
the coupling to dilaton for field strength includes the 
parameter $N$. 
Though there are classical solutions for 
particular values of $N$, the solutions of $N\ne 4$ models are  
no longer related to D-branes and M-branes. 
The dynamical solutions for $N=4$ 
were also developed independently 
in \cite{Binetruy:2007tu, Maeda:2009zi, Minamitsuji:2010kb}; 
the property of cosmological evolution had in essence been introduced 
earlier \cite{Kodama:2005fz, Kodama:2005cz}.
For any number of dimension, we will show that
the time-dependent solutions can be 
obtained for $N\ne 4$ by extending  
the ansatz. The case of $N\ne 4$ gives new intersecting brane solutions 
that all warp factors arising from field strengths can depend on time 
if the number $N$ has the appropriate values. There are also 
the dynamical intersecting solutions that one has $N=4$, the 
others have $N\ne 4$.  As a simple example, 
we will study the dynamical intersecting solution 
in a class of the six-dimensional Romans supergravity \cite{Romans:1985tw,Nunez:2001pt} 
with a vanishing cosmological constant. 
We will also see that the effect of a cosmological constant 
often changes the picture radically, 
in particular, triggering the accelerating expansion of the Universe. 
This can only happen 
when the scalar field vanishes,
since a nonzero scalar field is an
obstruction to accelerating expansion. 
Our results will also be interesting for cosmological 
applications of string theories. 

The dynamical solutions
in the six-dimensional Nishino-Salam-Sezgin (NSS) supergravity 
\cite{Nishino:1984gk, Salam:1984cj, Nishino:1986dc,
Gibbons:2003di,Aghababaie:2003ar} have been investigated
in~\cite{Maeda:1984gq,Maeda:1985es,Tolley:2006ht, 
Tolley:2007et,Minamitsuji:2010fp}, 
including applications to brane world models. 
A particular construction of dynamical solutions 
was discussed recently in this context and then 
applied to brane world models in \cite{Minamitsuji:2010fp} or 1-brane 
collision \cite{Maeda:2010aj}.
In the present paper, the dynamical 0-brane solution in the NSS model
will be derived as a special case
and used to study the possibility of brane collisions, 
which in the special case of $p$-branes
has been originally discussed in \cite{Gibbons:2005rt}.

If one drops the requirement of $N=4$ in the coupling to dilaton 
for field strengths, the solutions obtained 
in an Einstein-Maxwell model 
are a special case of a larger class of dynamical solutions  
that lead to de Sitter spacetime.
In Sec.~\ref{sec:cc}, we describe this larger class and 
apply it to construct brane world models
in the five-dimensional theory. 
In Sec.~\ref{sec:two}, we characterize the intersecting brane system 
that arises two kinds of form fields without the condition of $N=4$. 
In Sec.~\ref{sec:ni}, we perform explicit calculations 
illustrating how the dynamical solutions of $n$  
kinds of intersecting brane system 
 arise from the condition of $N\ne 4$. 
 These examples are inspired by and generalize an example
considered in Sec.~2 of \cite{Maeda:2009zi} as well as the detailed 
analysis of cosmological models in \cite{Minamitsuji:2010fp}.
Section \ref{sec:discussions} is devoted to concluding
remarks. 
\section{Dynamical solutions with a cosmological constant}
  \label{sec:cc}

\subsection{Theory}

We will start from the $D$-dimensional theory, 
for which the action in the Einstein frame contains the metric $g_{MN}$,
the scalar field $\phi$, the cosmological constant $\Lambda$,
and the antisymmetric tensor field of rank $(p+2)$, $F_{(p+2)}$
\Eq{
S=\frac{1}{2\kappa^2}\int \left[\left(R-2\e^{\alpha\phi}
\Lambda\right)\ast{\bf 1}_D
 -\frac{1}{2}\ast d\phi \wedge d\phi
 -\frac{1}{2\cdot (p+2)!}\e^{\epsilon c\phi}
 \ast F_{(p+2)}\wedge F_{(p+2)}\right],
\label{s:action:Eq}
}
where  $\alpha$ is constant, 
$\kappa^2$ is the $D$-dimensional gravitational constant,
$\ast$ is the Hodge operator in the $D$-dimensional spacetime,
$F_{(p+2)}$ is the $(p+2)$-form field strength,
and $c$, $\epsilon$ are constants given by
\Eqrsubl{s:parameters:Eq}{
c^2&=&N-\frac{2(p+1)(D-p-3)}{D-2},
   \label{s:c:Eq}\\
\epsilon&=&\left\{
\begin{array}{cc}
 +&~{\rm if}~~p-{\rm brane~is~electric}\\
 -&~~~{\rm if}~~p-{\rm brane~is~magnetic}\,.
\end{array} \right.
 \label{s:epsilon:Eq}
   }
Here, $N$ is a constant. 
The field strength $F_{(p+2)}$ is given by the
$(p+1)$-form gauge potential $A_{(p+1)}$
\Eq{F_{(p+2)}=dA_{(p+1)}.}

In this section,
we focus on dimensions of $D>2$.
In $D=10$ and $D=11$,
the cases of $\Lambda=0$ 
and $N=4$ of the theory (\ref{s:action:Eq}) correspond to
supergravities. 
The bosonic part of the action of $D=11$ supergravity includes only 4-form
 ($p=2$) without the scalar field,
since $c=0$ automatically. 
For $D=10$ and $N=4$, 
the constant $c$ is precisely the dilaton coupling for the 
Ramond-Ramond $(p+2)$-form in the type II supergravities. 
The dynamical solutions for the case of $N=4$
have been already discussed in 
\cite{Maeda:2010aj}. 
The bosonic part of the six-dimensional 
NSS model~\cite{Nishino:1984gk, Salam:1984cj, Nishino:1986dc}
is given by the expression (\ref{s:action:Eq})
with $\Lambda>0$. 
In this section, we will discuss the dynamical solution for $N\ne 4$.

After varying the action with respect to the metric, the scalar field,
and the $(p+1)$-form gauge field, the field equations are written by 
\Eqrsubl{s:equations:Eq}{
&&\hspace{-1cm}R_{MN}=\frac{2}{D-2}\e^{\alpha\phi}\Lambda g_{MN}
+\frac{1}{2}\pd_M\phi \pd_N \phi\nn\\
&&\hspace{-0.1cm}~~~~+\frac{1}{2\cdot (p+2)!}\e^{\epsilon c\phi}
\left[(p+2)F_{MA_2\cdots A_{p+2}} {F_N}^{A_2\cdots A_{p+2}}
-\frac{p+1}{D-2}g_{MN} F^2_{(p+2)}\right],
   \label{s:Einstein:Eq}\\
&&\hspace{-1cm}d\ast d\phi-\frac{\epsilon c}{2\cdot (p+2)!}
\e^{\epsilon c\phi}\ast F_{(p+2)}\wedge F_{(p+2)}
-2\alpha \e^{\alpha\phi}\Lambda\ast{\bf 1}_D=0,
   \label{s:scalar:Eq}\\
&&\hspace{-1cm}d\left[\e^{\epsilon c\phi}\ast F_{(p+2)}\right]=0.
   \label{s:gauge:Eq}
}

Now we assume that the $D$-dimensional metric
takes the form
\Eq{
ds^2=h^a(x, z)q_{\mu\nu}(\Xsp)dx^{\mu}dx^{\nu}
  +h^b(x, z)u_{ab}(\Zsp)dz^adz^b,
 \label{s:metric:Eq}
}
where $q_{\mu\nu}(\Xsp)$ denotes a $(p+1)$-dimensional metric which
depends only on the $(p+1)$-dimensional coordinates $x^{\mu}$,
and $u_{ab}(\Zsp)$ is the $(D-p-1)$-dimensional metric 
depending only on the $(D-p-1)$-dimensional coordinates $z^a$, and 
$\Xsp$ space represents the world volume directions,
while $\Zsp$ space does the space transverse to the $p$-brane.
The constants $a$ and $b$ are given by
\Eq{
a=-\frac{4(D-p-3)}{N(D-2)},~~~~b=\frac{4(p+1)}{N(D-2)}.
 \label{s:parameter:Eq}
}
The form of the metric (\ref{s:metric:Eq}) is
a straightforward generalization of the case of a static $p$-brane
system with a coupling of scalar field \cite{Lu:1995cs, Binetruy:2007tu}.
The scalar field $\phi$ and
the gauge field strength $F_{(p+2)}$ are assumed to be
\Eqrsubl{s:fields:Eq}{
&&\e^{\phi}=h^{2\epsilon c/N},
  \label{s:phi:Eq}\\
&&F_{(p+2)}=\frac{2}{\sqrt{N}}d(h^{-1})\wedge\Omega(\Xsp),
  \label{s:fp:Eq}
}
where $\Omega(\Xsp)$ is the volume $(p+1)$-form,
\Eq{
\Omega(\Xsp)=\sqrt{-q}\,dx^0\wedge dx^1\wedge \cdots \wedge
dx^p.
}
$q$ is the determinant of the metric $q_{\mu\nu}$.

\subsection{Asymptotically Milne solution}

Firstly, we consider the Einstein Eqs.~(\ref{s:Einstein:Eq}) with 
$c\neq 0$. 
We assume that the parameter $\alpha$ is given by 
\Eq{
\alpha=\left[-N+\frac{2(D-p-3)}{D-2}\right]\left(\epsilon c\right)^{-1}.
    \label{s:alpha:Eq}
}
Using the ansatz (\ref{s:metric:Eq}) and (\ref{s:fields:Eq}),
the Einstein equations are written by
\Eqrsubl{s:cEinstein:Eq}{
&&\hspace{-1.5cm}R_{\mu\nu}(\Xsp)-\frac{4}{N}h^{-1}D_{\mu}D_{\nu} h
+\frac{2}{N}\left(1-\frac{4}{N}\right)\pd_{\mu}\ln h\pd_{\nu}\ln h
-\frac{2}{D-2}\Lambda q_{\mu\nu}h^{-2}\nn\\
&&~~~-\frac{a}{2}q_{\mu\nu}\left[h^{-1}\lap_{\Xsp}h
-\left(1-\frac{4}{N}\right)q^{\rho\sigma}\pd_{\rho}\ln h\pd_{\sigma}
\ln h\right]-\frac{a}{2}q_{\mu\nu}h^{-4/N-1}\triangle_{\Zsp} h
=0,
 \label{s:cEinstein-mu:Eq}\\
&&\hspace{-1.5cm}h^{-1}\pd_{\mu}\pd_a h=0,
 \label{s:cEinstein-ma:Eq}\\
&&\hspace{-1.5cm}R_{ab}(\Zsp)-\frac{b}{2}h^{4/N} u_{ab}
\left[h^{-1}\triangle_{\Xsp} h-\left(1-\frac{4}{N}\right)
 q^{\rho\sigma}\pd_{\rho}\ln h\pd_{\sigma}\ln h\right]\nn\\
&&~~~-\frac{b}{2}u_{ab}h^{-1}\triangle_{\Zsp}h
-\frac{2}{D-2}\Lambda u_{ab}h^{-2+4/N}=0
 \label{s:cEinstein-ab:Eq},
}
where $D_{\mu}$ is the covariant derivative with respect to
the metric $q_{\mu\nu}$, $\triangle_{\Xsp}$ and $\triangle_{\Zsp}$ denote 
the Laplace operators on $\Xsp$ and $\Zsp$, respectively. 
Similarly, $R_{\mu\nu}(\Xsp)$ and $R_{ab}(\Zsp)$ are the Ricci tensors
associated with the metrics $q_{\mu\nu}$ and $u_{ab}$, respectively.
From Eq.~(\ref{s:cEinstein-ma:Eq}), 
the function $h$ have to be in the form
\Eq{
h(x, z)= h_0(x)+h_1(z).
  \label{s:warp:Eq}
}
With the form of the function $h$, the other components of the Einstein Eqs.
(\ref{s:cEinstein-mu:Eq}) and (\ref{s:cEinstein-ab:Eq}) are expressed as
\Eqrsubl{s:cEinstein2:Eq}{
&&R_{\mu\nu}(\Xsp)-\frac{4}{N}h^{-1}D_{\mu}D_{\nu} h_0
+\frac{2}{N}\left(1-\frac{4}{N}\right)h^{-2}\pd_{\mu}h_0\pd_{\nu}h_0
-\frac{2}{D-2}\Lambda q_{\mu\nu}h^{-2}\nn\\
&&~~~-\frac{a}{2}q_{\mu\nu}\left[h^{-1}\lap_{\Xsp}h_0
-\left(1-\frac{4}{N}\right)h^{-2}q^{\rho\sigma}\pd_{\rho}h_0\pd_{\sigma}h_0
\right]-\frac{a}{2}q_{\mu\nu}h^{-4/N-1}\triangle_{\Zsp} h_1=0,
   \label{s:cEinstein-mu2:Eq}\\
&&R_{ab}(\Zsp)-\frac{b}{2}h^{4/N} u_{ab}
\left[h^{-1}\triangle_{\Xsp} h_0-\left(1-\frac{4}{N}\right)
 h^{-2}q^{\rho\sigma}\pd_{\rho}h_0\pd_{\sigma}h_0\right]\nn\\
&&~~~-\frac{b}{2}u_{ab}h^{-1}\triangle_{\Zsp}h_1
-\frac{2}{D-2}\Lambda u_{ab}h^{-2+4/N}=0.
   \label{s:cEinstein-ab2:Eq}
   }

In terms of the assumption (\ref{s:fp:Eq}),
the Bianchi identity is automatically satisfied while 
the equation of motion for the gauge field~(\ref{s:gauge:Eq}) becomes
\Eq{
\lap_{\Zsp}h_1\,\Omega(\Zsp)=0,
 }
where we have used (\ref{s:warp:Eq}), and $\Omega(\Zsp)$ is defined by
\Eq{
\Omega(\Zsp)=\sqrt{u}\,dz^1\wedge\cdots\wedge dz^{D-p-1}.
}
Hence, the gauge field equation reduces to
\Eq{
\lap_{\Zsp}h_1=0.
   \label{s:h1:Eq}
}
We next consider the field equation of scalar field.
Substituting Eqs.~(\ref{s:fields:Eq}) and (\ref{s:warp:Eq}) into
Eq.~(\ref{s:scalar:Eq}), we find
\Eq{
\hspace{-0.2cm}
\frac{2}{N}\epsilon ch^{4/N-b}\left[h^{-1}\triangle_{\Xsp}h_0-
\left(1-\frac{4}{N}\right)
 h^{-2}q^{\rho\sigma}\pd_{\rho}h_0\pd_{\sigma}h_0
 +h^{-1-4/N}\triangle_{\Zsp}h_1\right]
-2\alpha h^{-2-a}\Lambda=0.
  \label{s:scalar2:Eq}
}
In terms of Eq.~(\ref{s:h1:Eq}), we are left with
\Eq{
\triangle_{\Xsp}h_0=0,~~~~\frac{1}{N}
\left(1-\frac{4}{N}\right)q^{\rho\sigma}
\pd_{\rho}h_0\pd_{\sigma}h_0+\left(\epsilon c\right)^{-1}\alpha \Lambda=0.
\label{s:scalar3:Eq}
}

Now we go back to the Einstein Eqs.~(\ref{s:cEinstein2:Eq}).
If $F_{(p+2)}=0$, the function $h_1$ becomes trivial.
If we set $F_{(p+2)}\ne 0$, the first term in 
Eq.~(\ref{s:cEinstein-mu2:Eq})
depends on only $x$ whereas the rest on both $x$ and $y$.
Then Eqs.~(\ref{s:cEinstein2:Eq}) together with (\ref{s:h1:Eq}) and
(\ref{s:scalar3:Eq}) lead to
\Eqrsubl{s:Einstein2:Eq}{
&&R_{\mu\nu}(\Xsp)=0,~~~~D_{\mu}D_{\nu}h_0=0,
   \label{s:Ricci-mn:Eq}\\
&&\frac{2}{N}\left(1-\frac{4}{N}\right)\pd_{\mu}h_0\pd_{\nu}h_0
+\frac{a}{2}\left(1-\frac{4}{N}\right)q_{\mu\nu}
q^{\rho\sigma}\pd_{\rho}h_0\pd_{\sigma}h_0
-\frac{2}{D-2}\Lambda q_{\mu\nu}=0,
   \label{s:Ricci-mn2:Eq}\\
&&R_{ab}(\Zsp)
+\frac{b}{2}\left(1-\frac{4}{N}\right)h^{-2+4/N} u_{ab}
q^{\rho\sigma}\pd_{\rho}h_0\pd_{\sigma}h_0
-\frac{2}{D-2}\Lambda u_{ab}h^{-2+4/N}=0.~~
   \label{s:Ricci-ab:Eq}
 }
The Eqs. (\ref{s:scalar3:Eq}) and (\ref{s:Ricci-mn2:Eq}) 
give $N=2$. Then, the Eqs. (\ref{s:Ricci-mn2:Eq}) and 
(\ref{s:Ricci-ab:Eq}) are written by
\Eqrsubl{s:Einstein3:Eq}{
&&\pd_{\mu}h_0\pd_{\nu}h_0=-\frac{4}{c^2(D-2)}\Lambda q_{\mu\nu},
   \label{s:Ricci-mn3:Eq}\\
&&R_{ab}(\Zsp)+\frac{4p}{c^2(D-2)}\Lambda u_{ab}=0,
   \label{s:Ricci-ab2:Eq}
}
respectively.
If one solves these Eqs.~(\ref{s:Einstein2:Eq}) with Eq. (\ref{s:h1:Eq}),
the solution of the present system is given by Eqs. (\ref{s:metric:Eq})
and (\ref{s:fields:Eq}) with (\ref{s:warp:Eq}).

For a nonvanishing cosmological constant, Eq.~(\ref{s:Ricci-ab:Eq})
implies that the $(D-p-1)$-dimensional space $\Zsp$ is an Einstein manifold.
The $(D-p-1)$-dimensional flat space is allowed only for $p=0$. 
Equation (\ref{s:Ricci-mn3:Eq}) implies that the 
$(p+1)$-dimensional metric $q_{\mu\nu}(\Xsp)$ is expressed as a 
product of two vectors. Hence, for $p\ne 0$, the determinant of the metric 
$q_{\mu\nu}(\Xsp)$ becomes zero, which is not permissible.
In the following, 
we will discuss the solution of $p=0$ case. 
We find that Eqs.~(\ref{s:h1:Eq}), (\ref{s:Einstein2:Eq}) and
 (\ref{s:Ricci-ab2:Eq}) reduce to
\Eqrsubl{s:fEinstein:Eq}{
&&h(t, z)=h_0(t)+h_1(z),~~~~h_0=At+B,~~~~\lap_{\Zsp}h_1=0,
   \label{s:fh:Eq}\\
&&R_{ab}(\Zsp)=0,
   \label{s:fRicci-ij:Eq}
 }
where $A$ is defined by $A\equiv\pm\sqrt{2\Lambda}$ 
and $B$ is constant parameter.
Thus,
there is no solution for $\Lambda<0$.
Upon setting
\Eq{
\quad u_{ab}=\delta_{ab}\,,
 \label{s:smetric:Eq}
 }
where $\delta_{ij}$ is
the $(D-1)$-dimensional Euclidean space metric,
the solution for $h$ is written explicitly as
\Eq{
h(t, z)=At+B+h_1(z)\,,
 \label{s:h2:Eq}
}
where the harmonic function $h_1$ is found to be
\Eqrsubl{s:harmonics:Eq}{
h_1(\,\vect{z})&=&\sum_{\ell=1}^{L}\frac{M_\ell}
{|\,\vect{z}-\,\vect{z}_\ell|^{D-3}}
~~~~~~{\rm for}~~D\neq 3\,,
\label{s:sh1:Eq}
\\
h_1(\,\vect{z})&=&\sum_{\ell=1}^{L}
M_\ell\, \ln |\,\vect{z}-\,\vect{z}_\ell|
~~~~{\rm for}~~D=3\,.
\label{s:sh2:Eq}
}
Here, $|\vect{z}-\,\vect{z}_\ell|
=\sqrt{\left(z^1-z^1_\ell\right)^2+\left(z^2-z^2_\ell\right)^2+\cdots+
\left(z^{D-1}-z^{D-1}_\ell\right)^2}$, and
$M_\ell~ (\ell=1\cdots L)$ are mass constants of 0-branes
located at $~\vect{z}_\ell$.
The behavior of the harmonic function $h_1$ is classified into two
classes depending on the dimensions $D$, i.e.
$D>3$, and $D=3$. 

For $D=3$, the harmonic function $h_1$ diverges
both at infinity and near 0-branes.
Moreover, because $h_1\rightarrow -\infty$,
there is no regular spacetime region near 0-branes.
Hence, such solutions are not physically relevant because 
the original theory is ill-defined.
In the following, we will focus on the case $D> 3$.

Assuming $\Lambda>0$, and
introducing a new time coordinate $\tau$ by 
\Eq{
\frac{\tau}{\tau_0}=
\big(At+B\big)^{1/(D-2)}\,,~~~~~
\tau_0=\frac{(D-2)}{A},
}
we find the $D$-dimensional metric (\ref{s:metric:Eq}) as
\Eq{
ds^2=
\left[1+\left(\frac{\tau}{\tau_0}\right)^{-(D-2)}h_1
\right]^{-\frac{2(D-3)}{D-2}}
\left[-d\tau^2+
\left\{1+\left(\frac{\tau}{\tau_0}\right)^{-(D-2)}h_1\right\}^2
\left(\frac{\tau}{\tau_0}\right)^2u_{ab}dz^adz^b\right].
 \label{s:s-metric:Eq}
 }
For $h_1\to 0$, the spacetime approaches
an isotropic and homogeneous universe,
whose scale factor is proportional to $\tau$,
i.e., the $D$-dimensional Milne universe.
This is realized in the limit $\tau\rightarrow\infty$, 
which is guaranteed by a scalar field with
the exponential potential.
The $D$-dimensional spacetime becomes inhomogeneous 
for $h_1\neq 0$. 
The power exponent of the
 scale factor is always larger than that in
the matter or radiation-dominated era. 
It is interesting to note that 
in the case of $D=6$, $p=0$
with $\Lambda>0$, Eq. 
(\ref{s:s-metric:Eq}) describes the cosmological solution 
in the NSS model with the vanishing 3-form field strength.
The late time evolution has a scaling behavior.
Note that the scaling solution in the NSS model
obtained in Ref. \cite{Tolley:2006ht}
has the similar time dependence,
although in this case
the 2-form field strength is magnetic. 

The $D$-dimensional spacetime is regular  
in the region of $h>0$,
but has curvature singularities where $h=0$,
since $\phi$ diverges there. 
The physical spacetime exists only inside the domain restricted by
\Eq{
h(t, \bm z) \equiv At+B+h_1(\,\vect{z})>0.
}

To see the detailed dynamics of spacetime, let us illustrate 
the case of two 0-branes,
which are sharing the same charge $M$ and 
located at $\,\vect{z}=(\pm L, 0,\cdots,0)$.
Here, we focus on the period of $t>0$.
For the period of $t<0$, the spacetime dynamics is obtained simply
by reversing evolution of the case of $t>0$.

In the case of $A>0$, for $t\geq0$ the metric is always regular.
The metric (\ref{s:metric:Eq}) 
implies that the transverse dimensions expand
 asymptotically as $\tilde \tau$, 
where $\tilde \tau$ is the proper time of the coordinate observer.
However, it is observer-dependent.
As we mentioned before, the $D$-dimensional spacetime becomes 
static near branes, and the spacetime approaches a Milne
universe in the far region ($|\,\vect{z}|\rightarrow \infty$),
which expands in all directions isotropically. 
Defining 
\Eq{
z_{\perp}=\sqrt{\left(z^2\right)^2+\cdots
+\left(z^{D-1}\right)^2}\,,
}
the proper distance at $z_{\perp}=0$
between two branes is given by
\begin{eqnarray}
d(t)&=&\int_{-L}^L dz^1 \left[At+{M\over |z^1+L|^{D-3}}
+{M\over |z^1-L|^{D-3}}
\right]^{1\over D-2}
\nonumber \\
&=&
\left({ML}\right)^{1\over D-2}
\int_{-1}^1 d\eta
\left[
\left({A L^{D-3}\over M}\right)\,
t+{1\over |\eta+1|^{D-3}}
+{1\over |\eta-1|^{D-3}}\right]^{1\over D-2}
\,,
\label{s:distance:Eq}
\end{eqnarray}
which is a monotonically increasing function of $t$.

Next, we discuss the case of $A<0$. 
All of the region of $(D-1)$-dimensional space is initially ($t=0$) regular
except at $z\rightarrow\infty$. 
The singular hypersurface erodes
the $\,\vect{z}$-coordinate region as time evolves. 
As a result, only the region near 0-branes remains regular.
When we consider this process on
the $(z^1, z_{\perp})$ plane, 
the singularity appears at infinity.
It eventually approaches 0-branes and finally
the regular spatial region splits
into two isolated throats surrounding each 0-brane.
The proper distance $d$ between two branes,
given by Eq. (\ref{s:distance:Eq}),
is now a monotonically decreasing function of $t$.
At a glance, it could 
realize brane collisions.
However, since a singularity appears between two branes
before the distance vanishes, 
a regular brane collision cannot be realized.

\subsection{Asymptotically de Sitter solution}

Next, we consider the solution with a trivial scalar field which is 
the case of $c=0$ and hence $\alpha=0$. 
The scalar field becomes constant because of the ansatz 
(\ref{s:scalar:Eq}), and the scalar field Eq. 
(\ref{s:scalar2:Eq}) is automatically satisfied. 
In terms of $c=0$, Eq.~(\ref{s:c:Eq}) give 
\Eq{
N=\frac{2(D-p-3)(p+1)}{D-2}.
   \label{sds:N:Eq}
}
In this case, the field equations are reduced to
\Eqrsubl{sds:equations:Eq}{
&&R_{\mu\nu}(\Xsp)=0,~~~~~R_{ab}(\Zsp)=0,\\
&&h(x, z)=h_0(x)+h_1(z),\\
&&D_{\mu}D_{\nu} h_0=0,~~~~
\pd_{\mu}h_0\pd_{\nu}h_0
+\frac{2(p+1)(D-p-3)^2}
      {(D-2)\big((1-p)D+p^2+4p-1\big)}
\Lambda q_{\mu\nu}=0,
  \label{sds:h0:Eq}\\
&&~~~\triangle_{\Zsp} h_1=0.
   \label{sds:h1:Eq}
   }
We will focus on the solution of $p=0$, 
since from Eq.~(\ref{sds:h0:Eq}) it turns out that 
the solution for $p\ne 0$ is not permissible.
Then Eq.~(\ref{sds:h0:Eq}) gives 
\Eq{
h_0=c_1t+c_2,
}
where $c_2$ is an integration constant and $c_1$ is given by
\Eq{
\label{c1}
c_1=\pm(D-3)\sqrt{\frac{2\Lambda}{(D-1)(D-2)}}.
}
Thus, there is no solution for $\Lambda<0$.
If the metric $u_{ab}(\Zsp)$ is assumed to be Eq.~(\ref{s:smetric:Eq}), 
the function $h_1$ is given by Eq.~(\ref{s:harmonics:Eq}). 
Now we introduce a new time coordinate $\tau$ by 
\Eq{
c_1\tau=\ln t,
}
where we have 
taken $c_1>0$ for simplicity. The $D$-dimensional metric 
(\ref{s:metric:Eq}) is then rewritten as
\Eq{
ds^2=-\left(1+c_1^{-1}\e^{-c_1\tau}h_1\right)^{-2}d\tau^2
+\left(1+c_1^{-1}\e^{-c_1\tau}h_1\right)^{2/(D-3)}
\left(c_1\e^{c_1\tau}\right)^{2/(D-3)}u_{ab}(\Zsp)dz^adz^b.
   \label{sds:ds:Eq}
}
Equation (\ref{sds:ds:Eq}) implies that the spacetime 
describes an isotropic and homogeneous universe 
if $h_1=0$.  In the limit when the terms with $h_1$ are negligible, 
which is realized in the limit $\tau\rightarrow\infty$ and for $c_1>0$, 
we find a $D$-dimensional de Sitter universe. 
The solution (\ref{sds:ds:Eq}) has been discussed by 
 \cite{Maki:1992tq} (see also \cite{Ivashchuk:1996zv}). 
Furthermore, for $D=4$, the solution is found by
Kastor and Traschen \cite{Kastor:1992nn}.

\subsection{Application to the brane world}

The asymptotically de Sitter solution in 
the case of $D=5$ 
is now applied
to construct a cosmological brane world.
We start from the general metric
\begin{eqnarray}
ds^2=-d(T,\xi)^2 dT^2
      +f(T, \xi)^2d\xi^2
      +a(T, \xi)^2d\Omega_{(3)}^2\,,
\label{metric_ge}
\end{eqnarray}
where $d\Omega_{(3)}^2$ denotes 
a unit 3-sphere.
$\xi$ and $T$
denote the dimensionless coordinates of the 
extra space and time.
For a given background spacetime,
applying the standard copy and paste method,
it is possible to construct a cosmological 3-brane world
embedded into a five-dimensional bulk
as in the Randall-Sundrum model \cite{rs}.
For simplicity, we impose the $Z_2$-symmetry across the brane world 
volume. 

A cosmological brane world evolves 
along a trajectory specified by an affine parameter $\tau$,
$(T,\xi)=(T(\tau),\xi(\tau))$.
The induced metric on the brane world is then
given by the closed Friedmann-Lemaitre-Robertson-Walker metric
with the scale factor $a$
\begin{eqnarray}
\label{ind}
ds_{\rm (ind)}^2
=- d\tau^2  +a(T(\tau), \xi(\tau))^2d\Omega_{(3)}^2\,,
\end{eqnarray}
where we imposed
\begin{eqnarray}
-d(T(\tau), \xi(\tau))^2 \dot{T}{}^2+f(T(\tau), \xi(\tau))^2\dot{\xi}{}^2=-1\,.
\label{normalization}
\end{eqnarray}
A dot denotes a derivative with respect to $\tau$,
which is interpreted as the cosmic proper time.

The trajectory of the brane world is through 
the junction conditions.
Here, we focus on the Israel conditions given by  
\Eqrsubl{junction}{
\frac{1}{2}\kappa^2{\bar \rho}&=&-
3\epsilon \Big\{
\frac{1}{f}
\frac{a_{,\xi}}{a}
\sqrt{1+f^2\dot{\xi}^2}
+\frac{f}{d}
\frac{a_{,T}}{a}
\dot{\xi}
  \Big\},
\label{junction1}
\\
\frac{1}{2}\kappa^2 {\bar p}&=&
\epsilon \Big\{
 \Big(
2\frac{a_{,\xi}}{a}
    +\frac{d_{,\xi}}{d}
\Big)
 \frac{1}{f}\sqrt{1+f^2\dot{\xi}^2}
+\frac{f_{,\xi}\dot{\xi}^2+\ddot{\xi}f}{\sqrt{1+f^2\dot{\xi}^2}}
+\frac{f}{d}\Big(
2\frac{a_{,T}}{a}
+ \frac{2f_{,T}}{f} \Big) 
\dot{\xi}
 \Big\}
\,,
\label{junction2}
}
where $\bar \rho$ and $\bar p$ represent 
the total energy density and pressure,
obtained by varying the brane world action.
From now on, we focus on the energy density equation, Eq. 
(\ref{junction1}).
Note that $\epsilon=+1$ denotes 
the normal vector pointing the direction of increasing $\xi$
(and $\epsilon=-1$ vice versa).

The derivative of the scale factor with respect to the cosmic proper time
is given by 
\begin{eqnarray}
\label{dota}
\dot{a}
= \alpha_{\xi} f\dot{\xi}
+\alpha_{T}\sqrt{1+f^2{\dot \xi}^2},
\end{eqnarray}
where we defined
\begin{eqnarray}
\alpha_\xi:=\frac{a_{,\xi}}{f},\quad
\alpha_T:=\frac{a_{,T}}{d}.
\end{eqnarray}
Replacing $\dot{\xi}$ with $\dot{a}$
through  Eq. (\ref{dota})
and squaring the energy density component of (\ref{junction1}),
we obtain the generalized Friedmann equation
for $a$.
In the $|\alpha_T|\gg |\alpha_\xi|$ limit, 
where the time dependence rules the spatial one,
the cosmological equation reduces to
\begin{eqnarray}
\frac{\dot{a}^2}{a^2}+\frac{1}{a^2}
\approx \frac{\kappa^4{\bar \rho}^2}
      {36}
+\frac{\alpha_T^2+1}{a^2}.
\label{dyn}
\end{eqnarray}
Similarly,
in the $|\alpha_T|\ll |\alpha_\xi|$ limit, 
where the spacetime is approximately static,
the cosmological equation reduces to
\begin{eqnarray}
\frac{\dot{a}^2}{a^2}+\frac{1}{a^2}
\approx \frac{\kappa^4{\bar \rho}^2}
      {36}
-\frac{\alpha_\xi^2-1}{a^2}.
\label{Teq}
\end{eqnarray}

\subsubsection{Brane world supported by the tension}

One possibility is to support the brane world 
by tension.
Decomposing $\bar \rho=\sigma+{\rho}$,
where $\sigma$ and $\rho$
denote the tension and matter energy density
localized on the brane, respectively,
we obtain
\bea
\frac{\dot{a}^2}{a^2}
+\frac{1}{a^2}
\approx
 \frac{1}{3}\Lambda_{\rm eff}
+\frac{\kappa_4^2}{3} \rho
+O(\rho^2).
\eea
Here, we assumed $\rho\ll\sigma$. 
The four-dimensional effective cosmological {constant}
(not exactly constant, of course)
and the gravitational constant are given by
\bea
\Lambda_{\rm eff}:=\frac{1}{12}\kappa^4\sigma^2
+\frac{3(\alpha_t^2+1)}{a^2},
\quad
\kappa_4^2:=\frac{1}{6}\kappa^4\sigma,
\eea
for $|\alpha_{T}|\gg|\alpha_{\xi}|$,
and
\bea
\Lambda_{\rm eff}:=\frac{1}{12}\kappa^4\sigma^2
-\frac{3(\alpha_\xi^2-1)}{a^2},
\quad
\kappa_4^2:=\frac{1}{6}\kappa^4\sigma,
\eea
for $|\alpha_{T}|\ll|\alpha_{\xi}|$.
$\Lambda_{\rm eff}$ is composed of
the tension part and the bulk part.
For $|\alpha_T|\gg |\alpha_{\xi}|$,
from Eq. (\ref{junction1})
assuming $\alpha_{T}>0$,
to obtain a positive gravitational constant,
namely, a positive tension,
we have to impose 
$\dot{\xi}>0$ for $\epsilon=-1$
and 
$\dot{\xi}<0$ for $\epsilon=+1$.
For $|\alpha_T|\ll |\alpha_{\xi}|$,
similarly from Eq. (\ref{junction1})
assuming $\alpha_{\xi}>0$,
to obtain a positive gravitational constant,
we have to impose 
$\epsilon=-1$.

\subsubsection{Brane world supported by the induced gravity}

The other possibility
is to support the brane world induced gravity term \cite{dgp}
\bea
\bar \rho={\rho}+ \mu^2 G_{(\rm ind)}{}^0{}_0
    = {\rho}-3\mu^2
\Big(\frac{\dot{a}^2}{a^2}+\frac{1}{a^2}\Big),
\eea
where $G_{(\rm ind)}{}_{\mu\nu}$ is
the Einstein tensor associated with the brane world metric
Eq. (\ref{ind}).
We will see that the parameter 
$\mu$ plays the role of the four-dimensional
Planck scale in the high density region.
We then obtain
for $|\alpha_{T}|\gg|\alpha_{\xi}|$, 
\bea
\Big(\frac{\dot{a}^2}{a^2}+\frac{1}{a^2}\Big)_{\pm}
=\frac{2}{\kappa^4\mu^4}
\Big[
1+\frac{\kappa^4\mu^2}{6}\rho
\pm \sqrt{1+\frac{\kappa^4 \mu^2}{3}\rho
 -\frac{\kappa^4\mu^4(\alpha_T^2+1)}{a^2}}
\Big],
\eea
and 
for $|\alpha_{T}|\ll|\alpha_{\xi}|$, 
\bea
\Big(\frac{\dot{a}^2}{a^2}+\frac{1}{a^2}\Big)_{\pm}
=\frac{2}{\kappa^4\mu^4}
\Big[
1+\frac{\kappa^4\mu^2}{6}\rho
\pm \sqrt{1+\frac{\kappa^4 \mu^2}{3}\rho 
+\frac{\kappa^4\mu^4(\alpha_\xi^2-1)}{a^2}}
\Big],
\eea
where in both cases
$(+)$ and $(-)$ denote two independent branches.
Here,
in the first limit,
if $\dot{\xi}>0$, 
we take $\epsilon=+1$ for the $(+)$-branch
and $\epsilon=-1$ for the $(-)$-branch (for $\dot{\xi}<0$ vice versa).
In the second limit,
we take $\epsilon=+1$ for the $(+)$-branch
and $\epsilon=-1$ for the (-)-branch.

Let us discuss the cosmological behaviors in
high and low energy density limits, respectively.
In both limits,
we recover the ordinary cosmological equation
in the high density region,
where the term linear in $\rho$ dominates others
\bea
\Big(\frac{\dot{a^2}}{a^2}+\frac{1}{a^2}
\Big)_{\pm}\approx \frac{1}{3\mu^2}\rho\,,
\label{recovery}
\eea
where clearly the four-dimensional gravitational 
constant is given by $\mu^{-1}$.
Thus, in this region,
the standard Friedmann equation
is recovered due to the induced gravity term.
On the other hand, in the low density region,
if $|\alpha_{T}|\gg|\alpha_{\xi}|$, 
\bea
\Big(\frac{\dot{a}^2}{a^2}+\frac{1}{a^2}\Big)_{\pm}
\simeq \frac{2}{\kappa^4\mu^4}
\Big(
1\pm \sqrt{1-\frac{\kappa^4\mu^4(\alpha_T^2+1)}{a^2}}
\Big),
\eea
and
if $|\alpha_{T}|\ll|\alpha_{\xi}|$, 
\bea
\Big(\frac{\dot{a}^2}{a^2}+\frac{1}{a^2}\Big)_{\pm}
\simeq \frac{2}{\kappa^4\mu^4}
\Big(
1\pm \sqrt{1+\frac{\kappa^4\mu^4(\alpha_\xi^2-1)}{a^2}}
\Big).
\eea

In the first limit,
for $\frac{\sqrt{\alpha_T^2+1}}{a}\ll\frac{1}{\kappa^2\mu^2}$,
$\big(\frac{\dot{a}}{a}\big)_{+}\approx \frac{2}{\kappa^2\mu^2}$
(ignoring the $\frac{1}{a^2}$ term compared to
the constant part)
and
$\big(\frac{\dot{a}}{a}\big)_{-}\approx 
\big|\frac{\alpha_T}{a}\big|$.
In the second limit,
for $\frac{|1-\alpha_\xi^2|^{\frac{1}{2}}}{a}\ll\frac{1}{\kappa^2\mu^2}$,
$\big(\frac{\dot{a}}{a}\big)_{+}\approx \frac{2}{\kappa^2\mu^2}$
and there is 
no regular behavior in the $(-)$-branch.
The (+)-branch
has the expansion rate of the self-accelerating solution 
given by Dvali, Gabadadze, and 
Porrati (DGP) \cite{dgp}.
However,
this branch is known to suffers a 
ghost instability \cite{ghost}.
On the other hand,
the (-)-branch does not contain any pathology.

\subsubsection{Brane world in the asymptotically de Sitter spacetime}

We apply our formulation
to the case of the asymptotically de Sitter solution
\eqref{sds:ds:Eq} in $D=5$.
Here, we have to assume that the harmonic function $h_1$ 
found in \eqref{s:sh1:Eq}
is given by the contribution of a single brane with a mass $M$.
Then, in $D=5$, the asymptotically de Sitter solution reduces to
\bea
ds^2
&=&
M\Big(1+\frac{\e^{- T}}{\xi^2}\Big)\e^{T} d\xi^2
 -c_1^{-2}\Big(1+\frac{\e^{- T}}{\xi^2}\Big)^{-2}dT^2
+M\xi^2\Big(1+\frac{\e^{- T}}{\xi^2}\Big)\e^{T}d\Omega_{(3)}^2,
\eea
where $c_1=\sqrt{\frac{2}{3}\Lambda}$ is given by Eq. \eqref{c1}
(we assume $c_1>0$).
The dimensionless coordinates run
$-\infty<T<\infty$ and $0<\xi<\infty$.
Since the combination in the round bracket is always positive,
no curvature singularity appears.
Comparing with \eqref{metric_ge},
$d$, $f$ and $a$ read
\bea
d:=c_1^{-1}\Big(1+\frac{\e^{-T}}{\xi^2}\Big)^{-1},\quad
f:=M^{\frac{1}{2}}\sqrt{1+\frac{e^{-T}}{\xi^2}}
 \e^{\frac{T}{2}},\quad
a:=M^{\frac{1}{2}}\xi
\sqrt{1+\frac{\e^{-T}}{\xi^2}}
 \e^{\frac{T}{2}},
\eea
and we find
\bea
\alpha_T
=\frac{\sqrt{c_1^2 M} \zeta(T,\xi)}{2}
\sqrt{1+\frac{1}{\zeta(T,\xi)^2}},\quad
\alpha_\xi
=\frac{\zeta(T,\xi)^2}
{1+\zeta(T,\xi)^2}<1,
\eea
where $\zeta(T,\xi):= \e^{\frac{T}{2}}\xi$.
We then define
\bea
F(\zeta):= \frac{\alpha_T}{\alpha_\xi}
=(c_1^2 M)^{\frac{1}{2}}\nu(\zeta),\quad
\nu(\zeta):=\frac{(1+\zeta^2)^{3/2}}{2\zeta^2}.
\eea
Here, $\nu(\zeta)$ takes minimum at $\zeta=\sqrt{2}$,
where $\nu(\sqrt{2})=\frac{3^{\frac{3}{2}}}{4}\simeq 1.299$.
Therefore as long as $c_1^2 M>\frac{16}{27}$,
we always obtain $\alpha_T>\alpha_{\xi}$.
In particular, for both limits of $\zeta\gg 1$ and $0<\zeta\ll 1$,
$\alpha_T\gg \alpha_{\xi}$, irrespective of $c_1^2 M$.
Thus,
the cosmological equation can be described
by Eq. (\ref{dyn}) with $\frac{\alpha_T}{a}=\frac{c_1}{2}$.

If the brane world is supported by the tension,
from \eqref{dyn}
the effective cosmological and gravitational constants
read
\bea
\Lambda_{\rm eff}
\approx
\frac{1}{12}\kappa^4\sigma^2+
\frac{3c_1^2}{4},
\quad
\kappa_4^2:=\frac{1}{6}\kappa^4\sigma.
\eea
To obtain a positive gravitational constant,
we impose $\epsilon=-1$ for $\dot{\xi}>0$
and $\epsilon=+1$ for $\dot{\xi}<0$.
In addition, 
the bulk volume is not finite in the direction of 
increasing $\xi$, as seen from
$
\sqrt{-g}=c_1 M^2 \xi^3 \e^{2T}\Big(1+\frac{\e^{-T}}{\xi^2}\Big).
$
Thus, to obtain the localized graviton on the brane world
at low energy,
$\xi$ must have an upper bound $0<\xi<\xi_0$,
where $\xi_0$ is the position of the brane world.
Thus, we also impose $\dot{\xi}>0$.
Then, the effective cosmology is the $\Lambda$CDM type one.

If the brane world is supported by the induced gravity,
we obtain
\bea
\Big(\frac{\dot{a}^2}{a^2}+\frac{1}{a^2}\Big)_{\pm}
=\frac{2}{\kappa^4\mu^4}
\Big[
1+\frac{\kappa^4\mu^2}{6}\rho
\pm \sqrt{1+\frac{\kappa^4 \mu^2}{3}\rho 
-\kappa^4\mu^4\Big(\frac{c_1^2}{4}+\frac{1}{a^2}\Big)}
\Big].
\eea
To ensure the regular cosmological behavior at the low energy
density, here we impose $c_1<\frac{2}{\kappa^2\mu^2}$. 
In the high density region, as shown in Eq.~\eqref{recovery},
the four-dimensional cosmological equation is recovered.
In the low density region,
\bea
\Big(\frac{\dot{a}^2}{a^2}+\frac{1}{a^2}\Big)_{\pm}
\simeq \frac{2}{\kappa^4\mu^4}
\Big(
1\pm 
\sqrt{1 -\kappa^4\mu^4\Big(\frac{c_1^2}{4}+\frac{1}{a^2}\Big)}
\Big).
\eea
If $c_1\ll\frac{2}{\kappa^2\mu^2}$,
$\big(\frac{\dot{a}}{a}\big)_{+}\approx \frac{2}{\kappa^2\mu^2}$
and
$\big(\frac{\dot{a}}{a}\big)_{-}\approx \frac{c_1}{2}$.
The latter shows that the healthy (-) branch,
as well as the (+)-branch can give
accelerating solutions in the later times.
The result is similar to a higher-dimensional extension of 
DGP \cite{sa}.
Although this property looks fascinating,
in order to explain the cosmic acceleration of today,
we have to require
that the bulk cosmological constant 
becomes very tiny as $\Lambda^{\frac{1}{2}}\simeq 10^{-42}{\rm GeV}$.
Therefore, for any reasonable choice of the five-dimensional
Planck scale, as ${\rm TeV}$ scale,
the huge fine-tuning for $\Lambda$ cannot be avoided.


\section{The intersection of two branes in $D$-dimensional theory}
  \label{sec:two}

\subsection{Theory}

In this section, we consider a $D$-dimensional theory 
composed of the metric $g_{MN}$,
scalar field $\phi$, and two antisymmetric tensor fields of rank $(p_r+2)$ 
and $(p_s+2)$: 
\Eqr{
S&=&\frac{1}{2\kappa^2}\int \left[R\ast{\bf 1}
 -\frac{1}{2}\ast d\phi \wedge d\phi
 -\frac{1}{2}\frac{1}{\left(p_r+2\right)!}
 \e^{\epsilon_rc_r\phi}\ast F_{(p_r+2)}\wedge F_{(p_r+2)}
 \right.\nn\\
 &&\left.
 -\frac{1}{2}\frac{1}{\left(p_s+2\right)!}
 \e^{\epsilon_sc_s\phi}\ast F_{(p_s+2)}\wedge F_{(p_s+2)}
 \right],
\label{it:action:Eq}
}
where $\kappa^2$ is the $D$-dimensional gravitational constant,  
$\ast$ is the Hodge operator in the $D$-dimensional spacetime, 
$F_{\left(p_r+2\right)}$ and $F_{\left(p_s+2\right)}$ 
are $\left(p_r+2\right)$-form, 
$\left(p_s+2\right)$-form field strengths, respectively, and 
$c_I$, $\epsilon_I~(I=r,~s)$ are constants given by 
\Eqrsubl{it:parameters:Eq}{
c_I^2&=&N_I-\frac{2(p_I+1)(D-p_I-3)}{D-2},
   \label{it:c:Eq}\\
\epsilon_I&=&\left\{
\begin{array}{cc}
 +&~{\rm if}~~p_I-{\rm brane~is~electric}\\
 -&~~~{\rm if}~~p_I-{\rm brane~is~magnetic}\,.
\end{array} \right.
 \label{it:epsilon:Eq}
   }
Here $N_I$ is constant. 
After varying the action
with respect to the metric, the scalar field, 
and the $\left(p_r+1\right)$-form and $\left(p_s+1\right)$-form 
gauge fields, the field equations are given by
\Eqrsubl{it:equations:Eq}{
&&R_{MN}=\frac{1}{2}\pd_M\phi \pd_N \phi
+\frac{1}{2}\frac{\e^{\epsilon_rc_r\phi}}
{\left(p_r+2\right)!}
\left[\left(p_r+2\right)
F_{MA_2\cdots A_{\left(p_r+2\right)}} 
{F_N}^{A_2\cdots A_{\left(p_r+2\right)}}
-\frac{p_r+1}{D-2} g_{MN} F^2_{\left(p_r+2\right)}\right]\nn\\
&&~~~~~~+\frac{1}{2}\frac{\e^{\epsilon_sc_s\phi}}
   {\left(p_s+2\right)!}
\left[\left(p_s+2\right)
F_{MA_2\cdots A_{\left(p_s+2\right)}} 
{F_N}^{A_2\cdots A_{\left(p_s+2\right)}}
-\frac{p_s+1}{D-2} g_{MN} F_{\left(p_s+2\right)}^2\right],
   \label{it:Einstein:Eq}\\
&&d\ast d\phi-\frac{1}{2}\frac{\epsilon_rc_r}
{\left(p_r+2\right)!}
\e^{\epsilon_rc_r\phi}\ast F_{\left(p_r+2\right)}\wedge F_{\left(p_r+2\right)}
-\frac{1}{2}\frac{\epsilon_sc_s}{\left(p_s+2\right)!}
\e^{\epsilon_sc_s\phi}\ast F_{\left(p_s+2\right)}\wedge
 F_{\left(p_s+2\right)}=0,
   \label{it:scalar:Eq}\\
&&d\left[\e^{\epsilon_rc_r\phi}\ast F_{\left(p_r+2\right)}\right]=0,
   \label{it:gauge-r:Eq}\\
&&d\left[\e^{\epsilon_sc_s\phi}\ast F_{\left(p_s+2\right)}\right]=0.
   \label{it:gauge-s:Eq}
}

We look for solutions whose $D$-dimensional metrics have the form
\Eqr{
ds^2&=&h^{a_r}_rh_s^{a_s}q_{\mu\nu}
(\Xsp)dx^{\mu}dx^{\nu}+h^{b_r}_rh_s^{a_s}\gamma_{ij}
(\Ysp_1)dy^idy^j\nn\\
&&+h^{a_r}_rh_s^{b_s}w_{mn}(\Ysp_2)dv^{m}dv^{n}
+h^{b_r}_rh_s^{b_s}u_{ab}(\Zsp)dz^adz^b, 
 \label{it:metric:Eq}
}
where $q_{\mu\nu}$ is a $(p+1)$-dimensional metric which
depends only on the $(p+1)$-dimensional coordinates $x^{\mu}$, 
$\gamma_{ij}$ is the $(p_s-p)$-dimensional metric which
depends only on the $(p_s-p)$-dimensional coordinates $y^i$, 
$w_{mn}$ is the $(p_r-p)$-dimensional metric which
depends only on the $(p_r-p)$-dimensional coordinates $v^m$
and finally $u_{ab}$ is the $(D+p-p_r-p_s-1)$-dimensional metric which
depends only on the $(D+p-p_r-p_s-1)$-dimensional coordinates $z^a$. 
The parameters $a_I~(I=r,~s)$ and $b_I~(I=r,~s)$ in the metric 
(\ref{it:metric:Eq}) are given by 
\Eq{
a_I=-\frac{4(D-p_I-3)}{N_I(D-2)},~~~~~b_I=\frac{4(p_I+1)}{N_I(D-2)}.
 \label{it:paremeter:Eq}
}

The $D$-dimensional metric (\ref{it:metric:Eq}) implies that 
the solutions are characterized by 
two functions, $h_r$ and $h_s$, which 
depend on the coordinates transverse to the brane as well as 
the world volume coordinate.
For the configurations of two branes, the powers of harmonic 
functions have to obey the intersection rule, 
and then split the coordinates in three parts. 
One is the overall world-volume coordinates, $\{x\}$, which are common 
to the two branes. The others are overall transverse coordinates,
$\{z\}$, and the 
relative transverse coordinates, $\{y\}$ and $\{v\}$,
which are transverse to only one 
of the two branes. 
The field equations of dynamical intersecting branes allow for the following 
three kinds of possibilities on $p_r$- and $p_s$-branes 
in $D$ dimensions 
\cite{Behrndt:1996pm, Bergshoeff:1996rn, Minamitsuji:2010kb}.\\
\begin{description}
\item

(I)Both $h_r$ and $h_s$ depend on the coordinates of 
overall transverse space: $h_r=h_r(x,z),\, h_s=h_s(x,z).$

\item

(II)Only $h_s$ depends on the coordinates of overall transverse space,
but the other $h_r$ does on the corresponding coordinates of 
relative transverse space: $h_r=h_r(x,y),\, h_s=h_s(x,z).$

\item

(III)Each of $h_r$ and $h_s$ depends on the corresponding 
coordinates of relative transverse space: $h_r=h_r(x,y),\, h_s=h_s(x,v).$

\end{description}

In the following, we discuss intersections where each participating 
brane corresponds to an independent harmonic function in the solution. 
We also derive the dynamical intersecting brane solution 
in $D$ dimensions obeying the above three conditions.  

\subsection{Case (I)}
  \label{sub:cI}
We first consider the case (I). Under our classification,
the $D$-dimensional metric ansatz Eq.~(\ref{it:metric:Eq})
now explicitly becomes 
\Eqr{
ds^2&=&h^{a_r}_r(x, z)h_s^{a_s}(x, z)q_{\mu\nu}
(\Xsp)dx^{\mu}dx^{\nu}+h^{b_r}_r(x, z)h_s^{a_s}(x, z)\gamma_{ij}
(\Ysp_1)dy^idy^j\nn\\
&&+h^{a_r}_r(x, z)h_s^{b_s}(x, z)w_{mn}(\Ysp_2)dv^{m}dv^{n}
+h^{b_r}_r(x, z)h_s^{b_s}(x, z)u_{ab}(\Zsp)dz^adz^b\,.  
 \label{c1:metric:Eq}
}
We also suppose that the 
scalar field $\phi$ and the gauge field strengths $F_{\left(p_r+2\right)}$, 
$F_{\left(p_s+2\right)}$ are written by
\Eqrsubl{c1:ansatz:Eq}{
\e^{\phi}&=&h_r^{2\epsilon_rc_r/N_r}\,
h_s^{2\epsilon_sc_s/N_s},
  \label{c1:scalar:Eq}\\
F_{\left(p_r+2\right)}&=&\frac{2}{\sqrt{N_r}}
d\left[h^{-1}_r(x, z)\right]\wedge\Omega(\Xsp)\wedge\Omega(\Ysp_2),
  \label{c1:gauge-r:Eq}\\
F_{\left(p_s+2\right)}&=&\frac{2}{\sqrt{N_s}}
d\left[h^{-1}_s(x, z)\right]\wedge\Omega(\Xsp)\wedge\Omega(\Ysp_1),
  \label{c1:gauge-s:Eq}
}
where $\Omega(\Xsp)$, $\Omega(\Ysp_1)$, and $\Omega(\Ysp_2)$ 
are the volume $(p+1)$-form, $(p_s-p)$-form, $(p_r-p)$-form, defined as
\Eqrsubl{c1:volume:Eq}{
\Omega(\Xsp)&=&\sqrt{-q}\,dx^0\wedge dx^1\wedge \cdots \wedge 
dx^p,\\
\Omega(\Ysp_1)&=&\sqrt{\gamma}\,dy^1\wedge dy^2\wedge \cdots \wedge 
dy^{p_s-p},\\
\Omega(\Ysp_2)&=&\sqrt{w}\,dv^1\wedge dv^2\wedge \cdots \wedge 
dv^{p_r-p}.
}
Here, $q$, $\gamma$, $w$ denote the determinant of the metric $q_{\mu\nu}$, 
$\gamma_{ij}$, $w_{mn}$, respectively.
Let us first consider the gauge field Eqs.~(\ref{it:gauge-r:Eq}), 
(\ref{it:gauge-s:Eq}).
From the assumptions (\ref{c1:gauge-r:Eq}) and  
(\ref{c1:gauge-s:Eq}), we get  
\Eqrsubl{c1:gauge2:Eq}{
&&d\left[h_s^{4\chi/N_s}\pd_a h_r\left(\ast_{\Zsp}dz^a\right)
\wedge\Omega(\Ysp_1)\right]=0,
  \label{c1:gauge2-r:Eq}\\
&&d\left[h_r^{4\chi/N_r}\pd_a h_s\left(\ast_{\Zsp}dz^a\right)
\wedge\Omega(\Ysp_2)\right]=0,
  \label{c1:gauge2-s:Eq}
 }
where $\ast_{\Zsp}$ denotes the Hodge operator on 
Z, and $\chi$ is defined by
\Eq{
\chi=
p+1-\frac{\left(p_r+1\right)\left(p_s+1\right)}{D-2}
+\frac{1}{2}\epsilon_r\epsilon_sc_rc_s\,. 
   \label{c1:chi:Eq}
}
Then, the Eq.~(\ref{c1:gauge2-r:Eq}) reduces to
\Eqrsubl{c1:gauge-r2:Eq}{
&&u^{ab}\pd_bh_s^{4\chi/N_s}\pd_a h_r
+h_s^{4\chi/N_s}\lap_{\Zsp}h_r=0,\\
&&\pd_{\mu}h_s^{4\chi/N_s}\pd_a h_r
+h_s^{4\chi/N_s}\pd_{\mu}\pd_a h_r=0, 
}
where $\triangle_{\Zsp}$ is the Laplace operators on the space of $\Zsp$. 
On the other hand, Eq.~(\ref{c1:gauge2-s:Eq}) leads to 
\Eqrsubl{c1:gauge-s2:Eq}{
&&u^{ab}\pd_bh_r^{4\chi/N_r}\pd_a h_s
+h_r^{4\chi/N_r}\lap_{\Zsp} h_s=0,\\
&&\pd_{\mu}h_r^{4\chi/N_r}\pd_a h_s
+h_r^{4\chi/N_r}\pd_{\mu}\pd_a h_s=0.
}
For $\chi=0$, the Eq.~(\ref{c1:gauge-r2:Eq}) reduces to
\Eq{
\lap_{\Zsp}h_r=0,
~~~\pd_{\mu}\pd_a h_r=0,
  \label{c1:gauge3:Eq}
}
and the Eq.~(\ref{c1:gauge-s2:Eq}) gives
\Eq{
\lap_{\Zsp}h_s=0,
~~~\pd_{\mu}\pd_a h_s=0.
  \label{c1:gauge4:Eq}
}
The relation $\chi=0$ is consistent with the intersection rule 
\cite{Tseytlin:1996bh, Argurio:1997gt, Argurio:1998cp, Ohta:1997gw, 
Aref'eva:1997nz, Ivashchuk:2001ra, Maeda:2009zi, Minamitsuji:2010kb}.

Next, we consider the Einstein Eq.~(\ref{it:Einstein:Eq}). 
Using the assumptions (\ref{it:metric:Eq}) and (\ref{c1:ansatz:Eq}), 
the Einstein equations are given by
\Eqrsubl{c1:cEinstein:Eq}{
&&\hspace{-0.3cm}R_{\mu\nu}(\Xsp)
-\frac{4}{N_r}h_r^{-1}D_{\mu}D_{\nu}h_r
-\frac{4}{N_s}h_s^{-1}D_{\mu}D_{\nu}h_s
+\frac{2}{N_r}\pd_{\mu}\ln h_r\left[\left(1-\frac{4}{N_r}\right)
\pd_{\nu}\ln h_r-\frac{4}{N_s}\pd_{\nu}\ln h_s\right]\nn\\
&&~~~~+\frac{2}{N_s}\pd_{\mu}\ln h_s\left[\left(1-\frac{4}{N_s}\right)
\pd_{\nu}\ln h_s-\frac{4}{N_r}\pd_{\nu}\ln h_r\right]\nn\\
&&~~~~-\frac{1}{2}q_{\mu\nu}h_r^{-4/N_r}h_s^{-4/N_s}
\left(a_rh_r^{-1}\lap_{\Zsp}h_r
+a_sh_s^{-1}\lap_{\Zsp}h_s\right)\nn\\
&&~~~~-\frac{1}{2}q_{\mu\nu}\left[a_r
h_r^{-1}\lap_{\Xsp}h_r-a_rq^{\rho\sigma}\pd_{\rho}\ln h_r
\left\{\left(1-\frac{4}{N_r}\right)
\pd_{\sigma}\ln h_r-\frac{4}{N_s}
\pd_{\sigma}\ln h_s\right\}\right.\nn\\
&&\left. ~~~~+a_sh_s^{-1}\lap_{\Xsp}h_s
-a_sq^{\rho\sigma}\pd_{\rho}\ln h_s\left\{\left(1-\frac{4}{N_s}\right)
\pd_{\sigma}\ln h_s-\frac{4}{N_r}\pd_{\sigma}\ln h_r\right\}\right]=0,
 \label{c1:cEinstein-mu:Eq}\\
&&h_r^{-1}\pd_{\mu}\pd_a h_r=0,
 \label{c1:cEinstein-mi:Eq}\\
&&h_s^{-1}\pd_{\mu}\pd_a h_s=0,
 \label{c1:cEinstein-mi2:Eq}\\
&&R_{ij}(\Ysp_1)-\frac{1}{2}h_r^{4/N_r}\gamma_{ij}\left[b_r
h_r^{-1}\lap_{\Xsp}h_r-b_rq^{\rho\sigma}\pd_{\rho}\ln h_r
\left\{\left(1-\frac{4}{N_r}\right)
\pd_{\sigma}\ln h_r-\frac{4}{N_s}
\pd_{\sigma}\ln h_s\right\}\right.\nn\\
&&\left. ~~~~+a_sh_s^{-1}\lap_{\Xsp}h_s
-a_sq^{\rho\sigma}\pd_{\rho}\ln h_s\left\{\left(1-\frac{4}{N_s}\right)
\pd_{\sigma}\ln h_s-\frac{4}{N_r}\pd_{\sigma}\ln h_r\right\}\right]\nn\\
&&~~~~-\frac{1}{2}\gamma_{ij}h_s^{-4/N_s}\left(b_rh_r^{-1}\lap_{\Zsp}h_r
+a_sh_s^{-1}\lap_{\Zsp}h_s\right)=0,
 \label{c1:cEinstein-ij:Eq}\\
&&R_{mn}(\Ysp_2)-\frac{1}{2}h_s^{4/N_s}w_{mn}\left[a_r
h_r^{-1}\lap_{\Xsp}h_r-a_rq^{\rho\sigma}\pd_{\rho}\ln h_r
\left\{\left(1-\frac{4}{N_r}\right)
\pd_{\sigma}\ln h_r-\frac{4}{N_s}
\pd_{\sigma}\ln h_s\right\}\right.\nn\\
&&\left. ~~~~+b_sh_s^{-1}\lap_{\Xsp}h_s
-b_sq^{\rho\sigma}\pd_{\rho}\ln h_s\left\{\left(1-\frac{4}{N_s}\right)
\pd_{\sigma}\ln h_s-\frac{4}{N_r}\pd_{\sigma}\ln h_r\right\}\right]\nn\\
&&~~~~-\frac{1}{2}w_{mn}h_r^{-4/N_r}\left(a_rh_r^{-1}\lap_{\Zsp}h_r
+b_sh_s^{-1}\lap_{\Zsp}h_s\right)=0,
 \label{c1:cEinstein-mn:Eq}\\
&&
R_{ab}(\Zsp)-\frac{1}{2}h_r^{4/N_r}h_s^{4/N_s}u_{ab}\left[b_r
h_r^{-1}\lap_{\Xsp}h_r-b_rq^{\rho\sigma}\pd_{\rho}\ln h_r
\left\{\left(1-\frac{4}{N_r}\right)
\pd_{\sigma}\ln h_r-\frac{4}{N_s}\pd_{\sigma}\ln h_s\right\}\right.\nn\\
&&\left. ~~~~+b_sh_s^{-1}\lap_{\Xsp}h_s
-b_sq^{\rho\sigma}\pd_{\rho}\ln h_s\left\{\left(1-\frac{4}{N_s}\right)
\pd_{\sigma}\ln h_s-\frac{4}{N_r}\pd_{\sigma}\ln h_r\right\}\right]\nn\\
&&~~~~-\frac{1}{2}u_{ab}\left(b_rh_r^{-1}\lap_{\Zsp}h_r
+b_sh_s^{-1}\lap_{\Zsp}h_s\right)=0,
  \label{c1:cEinstein-ab:Eq}
}
where we have used the intersection rule $\chi=0$, and
$D_{\mu}$ denotes the covariant derivative with respect to
the metric $q_{\mu\nu}$, $\triangle_{\Xsp}$ is
the Laplace operators on $\Xsp$ space, and
$R_{\mu\nu}(\Xsp)$, $R_{ij}(\Ysp_1)$, $R_{mn}(\Ysp_2)$,
and $R_{ab}(\Zsp)$ denote the Ricci tensors 
associated with the metrics $q_{\mu\nu}(\Xsp)$, $\gamma_{ij}(\Ysp_1)$,
$w_{mn}(\Ysp_2)$ and $u_{ab}(\Zsp)$, respectively.

From Eqs.~(\ref{c1:cEinstein-mi:Eq}) and (\ref{c1:cEinstein-mi2:Eq}), 
we find that the warp factors $h_r$ and $h_s$ must 
take the form
\Eq{
h_r(x, z)= h_0(x)+h_1(z),~~~~h_s(x, z)= k_0(x)+k_1(z).
  \label{c1:warp:Eq}
}
In terms of this form of $h_r$ and $h_s$, the other components of
the Einstein Eqs.~(\ref{c1:cEinstein:Eq}) are replaced as
\Eqrsubl{c1:c2Einstein:Eq}{
&&\hspace{-0.3cm}R_{\mu\nu}(\Xsp)
-\frac{4}{N_r}h_r^{-1}D_{\mu}D_{\nu}h_0
-\frac{4}{N_s}h_s^{-1}D_{\mu}D_{\nu}k_0
+\frac{2}{N_r}\pd_{\mu}\ln h_r\left[\left(1-\frac{4}{N_r}\right)
\pd_{\nu}\ln h_r-\frac{4}{N_s}\pd_{\nu}\ln h_s\right]\nn\\
&&~~~~+\frac{2}{N_s}\pd_{\mu}\ln h_s\left[\left(1-\frac{4}{N_s}\right)
\pd_{\nu}\ln h_s-\frac{4}{N_r}\pd_{\nu}\ln h_r\right]\nn\\
&&~~~~-\frac{1}{2}h_r^{-4/N_r}h_s^{-4/N_s}
q_{\mu\nu}\left(a_rh_r^{-1}\lap_{\Zsp}h_1
+a_sh_s^{-1}\lap_{\Zsp}k_1\right)\nn\\
&&~~~~-\frac{1}{2}q_{\mu\nu}\left[a_r
h_r^{-1}\lap_{\Xsp}h_0-a_rq^{\rho\sigma}\pd_{\rho}\ln h_r
\left\{\left(1-\frac{4}{N_r}\right)
\pd_{\sigma}\ln h_r-\frac{4}{N_s}
\pd_{\sigma}\ln h_s\right\}\right.\nn\\
&&\left. ~~~~+a_sh_s^{-1}\lap_{\Xsp}k_0
-a_sq^{\rho\sigma}\pd_{\rho}\ln h_s\left\{\left(1-\frac{4}{N_s}\right)
\pd_{\sigma}\ln h_s-\frac{4}{N_r}\pd_{\sigma}\ln h_r\right\}\right]=0,
 \label{c1:c2Einstein-mu:Eq}\\
&&R_{ij}(\Ysp_1)-\frac{1}{2}h_r^{4/N_r}\gamma_{ij}\left[b_r
h_r^{-1}\lap_{\Xsp}h_0-b_r
q^{\rho\sigma}\pd_{\rho}\ln h_r
\left\{\left(1-\frac{4}{N_r}\right)
\pd_{\sigma}\ln h_r-\frac{4}{N_s}
\pd_{\sigma}\ln h_s\right\}\right.\nn\\
&&\left. ~~~~+a_sh_s^{-1}\lap_{\Xsp}k_0
-a_s q^{\rho\sigma}\pd_{\rho}\ln h_s\left\{\left(1-\frac{4}{N_s}\right)
\pd_{\sigma}\ln h_s-\frac{4}{N_r}\pd_{\sigma}\ln h_r\right\}\right]\nn\\
&&~~~~-\frac{1}{2}\gamma_{ij}h_s^{-4/N_s}\left(b_rh_r^{-1}\lap_{\Zsp}h_1
+a_sh_s^{-1}\lap_{\Zsp}k_1\right)=0,
 \label{c1:c2Einstein-ij:Eq}\\
&&R_{mn}(\Ysp_2)-\frac{1}{2}h_s^{4/N_s}w_{mn}\left[a_r
h_r^{-1}\lap_{\Xsp}h_0-a_rq^{\rho\sigma}\pd_{\rho}\ln h_r
\left\{\left(1-\frac{4}{N_r}\right)
\pd_{\sigma}\ln h_r-\frac{4}{N_s}
\pd_{\sigma}\ln h_s\right\}\right.\nn\\
&&\left. ~~~~+b_sh_s^{-1}\lap_{\Xsp}k_0
-b_sq^{\rho\sigma}\pd_{\rho}\ln h_s\left\{\left(1-\frac{4}{N_s}\right)
\pd_{\sigma}\ln h_s-\frac{4}{N_r}\pd_{\sigma}\ln h_r\right\}\right]\nn\\
&&~~~~-\frac{1}{2}w_{mn}h_r^{-4/N_r}\left(a_rh_r^{-1}\lap_{\Zsp}h_1
+b_sh_s^{-1}\lap_{\Zsp}k_1\right)=0,
 \label{c1:c2Einstein-mn:Eq}\\
&&R_{ab}(\Zsp)-\frac{1}{2}h_r^{4/N_r}h_s^{4/N_s}u_{ab}\left[b_r
h_r^{-1}\lap_{\Xsp}h_0-b_r
q^{\rho\sigma}\pd_{\rho}\ln h_r\left\{\left(1-\frac{4}{N_r}\right)
\pd_{\sigma}\ln h_r-\frac{4}{N_s}\pd_{\sigma}\ln h_s\right\}\right.\nn\\
&&\left. ~~~~+b_sh_s^{-1}\lap_{\Xsp}k_0
-b_sq^{\rho\sigma}\pd_{\rho}\ln h_s\left\{\left(1-\frac{4}{N_s}\right)
\pd_{\sigma}\ln h_s-\frac{4}{N_r}\pd_{\sigma}\ln h_r\right\}\right]\nn\\
&&~~~~-\frac{1}{2}u_{ab}\left(b_rh_r^{-1}\lap_{\Zsp}h_1
+b_sh_s^{-1}\lap_{\Zsp}k_1\right)=0.
  \label{c1:c2Einstein-ab:Eq}
}
Finally we have to consider the scalar field equation.
Substituting Eqs.~(\ref{c1:ansatz:Eq}), (\ref{c1:warp:Eq}) and
the intersection rule $\chi=0$ into Eq.~(\ref{it:scalar:Eq}), we find
\Eqr{
&&\frac{\epsilon_rc_r}{N_r}h_r^{4/N_r}h_s^{4/N_s}\left[
h_r^{-1}\lap_{\Xsp}h_0-
q^{\rho\sigma}\pd_{\rho}\ln h_r\left\{\left(1-\frac{4}{N_r}\right)
\pd_{\sigma}\ln h_r-\frac{4}{N_s}\pd_{\sigma}\ln h_s\right\}\right]\nn\\
&&~~~~~+\frac{\epsilon_sc_s}{N_s}h_r^{4/N_r}h_s^{4/N_s}\left[
h_s^{-1}\lap_{\Xsp}k_0-
q^{\rho\sigma}\pd_{\rho}\ln h_s\left\{\left(1-\frac{4}{N_s}\right)
\pd_{\sigma}\ln h_s-\frac{4}{N_r}\pd_{\sigma}\ln h_r\right\}\right]\nn\\
&&~~~~~+\frac{\epsilon_rc_r}{N_r}h_r^{-1}\lap_{\Zsp}h_1
+\frac{\epsilon_sc_s}{N_s}h_s^{-1}\lap_{\Zsp}k_1=0.
  \label{c1:scalar2:Eq}
}
Then, the warp factors $h_r$ and $h_s$ should obey the equations
\Eqrsubl{c1:scalar solution:Eq}{
&&\triangle_{\Xsp}h_0-q^{\rho\sigma}\pd_{\rho}h_0
\left[\left(1-\frac{4}{N_r}\right)
\pd_{\sigma}\ln h_r-\frac{4}{N_s}\pd_{\sigma}\ln h_s\right]=0,
~~ \triangle_{\Zsp}h_1=0,
   \label{c1:scalar solution1:Eq}\\
&&\triangle_{\Xsp}k_0-q^{\rho\sigma}\pd_{\rho}k_0
\left[\left(1-\frac{4}{N_s}\right)
\pd_{\sigma}\ln h_s-\frac{4}{N_r}\pd_{\sigma}\ln h_r\right]=0,
~~ \triangle_{\Zsp}k_1=0.
   \label{c1:scalar solution2:Eq}
}

Combining these, we find that these field equations reduce to
\Eqrsubl{c1:Einstein:Eq}{
&&R_{\mu\nu}(\Xsp)=0,~~~~R_{ij}(\Ysp_1)=0,~~~~
R_{mn}(\Ysp_2)=0,~~~~R_{ab}(\Zsp)=0,
   \label{c1:Ricci:Eq}\\
&&h_r=h_0(x)+h_1(z),~~~~h_s=k_0(x)+k_1(z),
   \label{c1:h:Eq}\\
&&D_{\mu}D_{\nu}h_0=0, ~~~\pd_{\mu}h_0\left[\left(1-\frac{4}{N_r}\right)
\pd_{\nu}\ln h_r-\frac{4}{N_s}\pd_{\nu}\ln h_s\right]=0,
~~~\triangle_{\Zsp}h_1=0,
   \label{c1:warp2:Eq}\\
&&D_{\mu}D_{\nu}k_0=0,~~~\pd_{\mu}k_0\left[\left(1-\frac{4}{N_s}\right)
\pd_{\nu}\ln h_s-\frac{4}{N_r}\pd_{\nu}\ln h_r\right]=0,
~~~\triangle_{\Zsp}k_1=0.
   \label{c1:warp3:Eq}
 }
Upon setting $F_{\left(p_r+2\right)}=0$ and $F_{\left(p_s+2\right)}=0$,
the functions $h_1$ and $k_1$ become trivial, 
and the $D$-dimensional spacetime is no longer warped~
\cite{Kodama:2005fz, Kodama:2005cz}.

\subsubsection{The case of $\frac{1}{N_r}+\frac{1}{N_s}=\frac{1}{4}$}

To see the solutions more explicitly, let us consider the case
\Eq{
q_{\mu\nu}=\eta_{\mu\nu}\,,~~~\gamma_{ij}=\delta_{ij}\,,~~~
w_{mn}=\delta_{mn}\,,~~~u_{ab}=\delta_{ab}\,,~~~~
\frac{1}{N_r}+\frac{1}{N_s}=\frac{1}{4},~~~~h_r=h_s,
 \label{c1:flat:Eq}
 }
where $\eta_{\mu\nu}$ is the $(p+1)$-dimensional
Minkowski metric and $\delta_{ij}$, $\delta_{mn}$, $\delta_{ab}$ are
the $(p_s-p)$-, $(p_r-p)$- and $(D+p-p_r-p_s-1)$-dimensional Euclidean metrics,
respectively. This physically means that both branes have the same 
total amount of charge. 
The solution for $h_r$ and $h_s$ can be
written explicitly by
\Eqrsubl{c1:solution:Eq}{
h_r(x, z)&=&A_{\mu}x^{\mu}+B
+\sum_{\ell}\frac{M_\ell}{|\bm z-\bm z_\ell|^{D+p-p_r-p_s-3}},
 \label{c1:solution-r:Eq}\\
h_s(x, z)&=&h_r(x, z),
 \label{c1:solution-s:Eq}
}
where $A_{\mu}$, $B$, $C$, $M_\ell$ and $M_c$ denote constant parameters,
and $\bm z_\ell$ and $\bm z_c$ are
constant vectors representing the positions of the branes.
Since the functions coincide, the locations of the 
brane will also coincide. 

Let us consider the intersection rule in the $D$-dimensional
theory. For $p_r=p_s$ and 
$N_r=N_s=8$, 
the intersection rule $\chi=0$ leads to
\Eq{
p=p_r-4.
    \label{c1:chi2:Eq}
}
Then, we get the intersection involving two $p_r$-brane
\Eq{
p_r\cap p_r=p_r-4.
   \label{c1:int:Eq}
}
Equation~\eqref{c1:int:Eq} tells us that the numbers of
intersection for $p_r<4$ are negative,
which means that there is no intersecting solution of these brane systems. 
Since $p$ is positive or zero, 
the number of the total dimension must be $D\ge 10$.

\subsubsection{The case of $\frac{1}{N_r}+\frac{1}{N_s}\neq \frac{1}{4}$}

For the case
$\frac{1}{N_r}+\frac{1}{N_s}\neq \frac{1}{4}$, 
the field equations can be satisfied 
only if there is only one function $h_I (I=r~{\rm or}~s)$ 
depending on both $z^a$ 
and $x^{\mu}$, and other functions are either dependent
on $z^a$ or constant.

In the case of $\frac{1}{N_r}+\frac{1}{N_s}\neq \frac{1}{4}$,
if $\partial_{\mu}k_0=0$,
from Eq. (\ref{c1:scalar solution:Eq})
we obtain $(1-\frac{4}{N_r})\times
q^{\rho\sigma}\partial_{\rho}h_0 \partial_\sigma h_0=0$
with $\lap_{\Xsp} h_0=0$.
Thus, in this case
it is clear
that there is no solution for $h_0(x)$
such as $\partial_{\mu}h_0\neq 0$
unless $N_r=4$. 
Then we set
\Eq{
q_{\mu\nu}=\eta_{\mu\nu}\,,~~~\gamma_{ij}=\delta_{ij}\,,~~~
w_{mn}=\delta_{mn}\,,~~~u_{ab}=\delta_{ab}\,,~~~N_r=4\,,
 \label{c1:flat2:Eq}
 }
where $\eta_{\mu\nu}$ is the $(p+1)$-dimensional
Minkowski metric and $\delta_{ij}$, $\delta_{mn}$, $\delta_{ab}$ are
the $(p_s-p)$-, $(p_r-p)$- and $(D+p-p_r-p_s-1)$-dimensional Euclidean metrics,
respectively. For $\pd_{\mu}h_s=0$, the solution for $h_r$ and $h_s$ can be
obtained explicitly as
\Eqrsubl{c1:solution2:Eq}{
h_r(x, z)&=&A_{\mu}x^{\mu}+B
+\sum_{\ell}\frac{M_\ell}{|\bm z-\bm z_\ell|^{D+p-p_r-p_s-3}},
 \label{c1:solution-r2:Eq}\\
h_s(z)&=&C+\sum_{c}\frac{M_c}{|\bm z-\bm z_c|^{D+p-p_r-p_s-3}},
 \label{c1:solution-s2:Eq}
}
where $A_{\mu}$, $B$, $C$, $M_\ell$ and $M_c$ are constant parameters,
and $\bm z_\ell$ and
$\bm z_c$ denote constant vectors representing the positions of the branes.
Thus, in our time-dependent generalization of intersecting brane
solutions, only D-, NS-branes as well as M-branes can be time dependent 
in the $D=11$ and $D=10$ supergravities. 

Note that the conditions on the intersection $\chi=0$
(see Eq.~(\ref{c1:chi:Eq})) have now potentially more solutions,
since $p$ can now take also the value $p=-1$ 
(thus defining an intersection on a point in Euclidean space). 
There is also a single brane solution which only exists in the Euclidean
formulation is the (-1)-brane, or D-instanton for $\partial_{\mu}k_0=0$.
The intersections for the (-1)-brane (D-instantons) in $N_I=4$ are given 
by~\cite{Argurio:1997gt,Ohta:1997gw,Minamitsuji:2010kb}
\Eqrsubl{c1:rule:Eq}{
&&{\rm D}(-1)\cap {\rm D}p = \frac{1}{2}\left(p-5\right),\\
   \label{c1:int D:Eq}
&&{\rm D}(-1)\cap {\rm F1}=0.
   \label{c1:int F:Eq}
}

We then consider the NSS model among the theories of $D=6$. 
The couplings of the 3-form ($p_r=1$)
and the 2-form ($p_s=0$) field strengths to the dilaton 
are given by
$\epsilon_rc_r=-\sqrt{2}$, $\epsilon_sc_s=-\frac{1}{\sqrt{2}}$,
respectively. 
From Eq.~(\ref{it:c:Eq}), this case 
is realized by choosing $N_r=4$ and $N_s=2$. 
However, the number of the intersections dimensions 
is $-1$, according to the intersection rule $\chi=0$. 
Though meaningless in ordinary spacetime, 
these configurations are  
relevant in the Euclidean space, 
for instance representing instantons. 

Similarly, for the ${\cal N}=4^{g}$ class 
of the six-dimensional Romans theory \cite{Romans:1985tw},
following the classification in Ref. \cite{Nunez:2001pt},
the coupling  
of the 3-form and of the 2-form field strengths to the dilaton are  
given by $\epsilon_rc_r=-\sqrt{2}$, $\epsilon_sc_s=1/\sqrt{2}$, respectively.  
The number of the intersecting dimensions 
is zero from the intersection rule. On the other hand,
for the ${\cal N}={\tilde 4}^{g}$ class,
the number of the intersecting dimension becomes $-1$ as for the NSS model
and therefore the solution is classically meaningless.

\subsection{Case (II)}
  \label{sub:cII}

Let us next consider the case (II).
The $D$-dimensional metric ansatz (\ref{it:metric:Eq}) reduces to
\Eqr{
ds^2&=&h^{a_r}_r(x, y)h_s^{a_s}(x, z)q_{\mu\nu}
(\Xsp)dx^{\mu}dx^{\nu}+h^{b_r}_r(x, y)h_s^{a_s}(x, z)\gamma_{ij}
(\Ysp_1)dy^idy^j\nn\\
&&+h^{a_r}_r(x, y)h_s^{b_s}(x, z)w_{mn}(\Ysp_2)dv^{m}dv^{n}
+h^{b_r}_r(x, y)h_s^{b_s}(x, z)u_{ab}(\Zsp)dz^adz^b.
 \label{c2:metric:Eq}
}
The scalar field $\phi$ and the gauge field strengths are also assumed to 
be 
\Eqrsubl{c2:ansatz:Eq}{
\e^{\phi}&=&h_r^{2\epsilon_rc_r/N_r}\,
h_s^{2\epsilon_sc_s/N_s},
  \label{c2:scalar:Eq}\\
F_{\left(p_r+2\right)}&=&\frac{2}{\sqrt{N_r}}
d\left[h^{-1}_r(x, y)\right]\wedge\Omega(\Xsp)
\wedge\Omega(\Ysp_2),
  \label{c2:gauge-r:Eq}\\
F_{\left(p_s+2\right)}&=&\frac{2}{\sqrt{N_s}}
d\left[h^{-1}_s(x, z)\right]\wedge\Omega(\Xsp)
\wedge\Omega(\Ysp_1),
  \label{c2:gauge-s:Eq}
}
where $\Omega(\Xsp)$, $\Omega(\Ysp_1)$, and $\Omega(\Ysp_2)$
are given by (\ref{c1:volume:Eq}).
Under the same procedure as in Sec.~\ref{sub:cI}, we can find 
the intersection rule $\chi=0$ from the field equations.
Setting $\chi=0$, it is easy to show that the field equations lead to 
\Eqrsubl{c2:solution of Einstein:Eq}{
&&R_{\mu\nu}(\Xsp)=0,~~~~R_{ij}(\Ysp_1)=0,~~~~
R_{mn}(\Ysp_2)=0,~~~~R_{ab}(\Zsp)=0,
   \label{c2:Ricci:Eq}\\
&&h_r=h_0(x)+h_1(y),~~~~h_s=k_0(x)+k_1(z),
   \label{c2:h:Eq}\\
&&D_{\mu}D_{\nu}h_0=0, ~~~\pd_{\mu}h_0\left[\left(1-\frac{4}{N_r}\right)
\pd_{\nu}\ln h_r-\frac{4}{N_s}\pd_{\nu}\ln h_s\right]=0,
~~~\triangle_{\Ysp_1}h_1=0,
   \label{c2:warp:Eq}\\
&&D_{\mu}D_{\nu}k_0=0,~~~\pd_{\mu}k_0\left[\left(1-\frac{4}{N_s}\right)
\pd_{\nu}\ln h_s-\frac{4}{N_r}\pd_{\nu}\ln h_r\right]=0,
~~~\triangle_{\Zsp}k_1=0\,,
   \label{c2:warp2:Eq}
 }
where $\triangle_{\Ysp_1}$ is the Laplace
operators on the space of $\Ysp_1$.
If we set $F_{\left(p_r+2\right)}\ne 0$ and $F_{\left(p_s+2\right)}\ne 0$,
the functions $h_1$ and $k_1$ are nontrivial.
There is no dynamical solution for 
$\frac{1}{N_r}+\frac{1}{N_s}= \frac{1}{4}$ because we can not 
take both functions to be equal.

In the case of $\frac{1}{N_r}+\frac{1}{N_s}\neq \frac{1}{4}$,
as in the case (I), if $\partial_{\mu}k_0=0$
there is no solution for $h_0(x)$
such as $\partial_{\mu}h_0\neq 0$ unless $N_r=4$.
We discuss the following case in more detail:
\Eq{
q_{\mu\nu}=\eta_{\mu\nu}\,,~~~\gamma_{ij}=\delta_{ij}\,,~~~
w_{mn}=\delta_{mn}\,,~~~u_{ab}=\delta_{ab}\,,~~~~N_r=4\,,
 \label{c2:flat:Eq}
 }
where $\eta_{\mu\nu}$ denotes the $(p+1)$-dimensional
Minkowski metric and $\delta_{ij}$, $\delta_{mn}$, $\delta_{ab}$ are
the $(p_s-p)$-, $(p_r-p)$-, and $(D+p-p_r-p_s-1)$-dimensional 
Euclidean metrics, respectively. 
For $\pd_{\mu}h_s=0$, the functions $h_r$ and
$h_s$ can be expressed explicitly as
\Eqrsubl{c2:hrs:Eq}{
h_r(x, y)&=&A_{\mu}x^{\mu}+B
+\sum_{\ell}\frac{M_\ell}{|\bm y-\bm y_\ell|^{p_s-p-2}},
 \label{c2:hr:Eq}\\
h_s(z)&=&C+\sum_{c}\frac{M_c}{|\bm z-\bm z_c|^{D+p-p_r-p_s-3}},
 \label{c2:hs:Eq}
}
where $A_{\mu}$, $B$, $C$, $\bm y_\ell$, $\bm z_c$, $M_\ell$
and $M_c$ denote constant parameters.
Thus, in our time-dependent generalization of intersecting brane
solutions, only D-, NS-branes as well as M-branes can be time dependent 
in the $D=11$ and $D=10$ supergravities. 

\subsection{Case (III)}
  \label{sub:cIII}

Finally, we discuss the case (III).
From the $D$-dimensional metric ansatz (\ref{it:metric:Eq}), we find
\Eqr{
ds^2&=&h^{a_r}_r(x, y)h_s^{a_s}(x, v)q_{\mu\nu}
(\Xsp)dx^{\mu}dx^{\nu}+h^{b_r}_r(x, y)h_s^{a_s}(x, v)\gamma_{ij}
(\Ysp_1)dy^idy^j\nn\\
&&+h^{a_r}_r(x, y)h_s^{b_s}(x, v)w_{mn}(\Ysp_2)dv^{m}dv^{n}
+h^{b_r}_r(x, y)h_s^{b_s}(x, v)u_{ab}(\Zsp)dz^adz^b.
 \label{c3:metric:Eq}
}
We also take the following ansatz for the scalar field $\phi$ and
the gauge field strengths: 
\Eqrsubl{c3:ansatz:Eq}{
\e^{\phi}&=&h_r^{2\epsilon_rc_r/N_r}\,
h_s^{2\epsilon_sc_s/N_s},
  \label{c3:scalar:Eq}\\
F_{\left(p_r+2\right)}&=&\frac{2}{\sqrt{N_r}}
h_s^{4/N_s}\,d\left[h^{-1}_r(x, y)\right]\wedge\Omega(\Xsp)
\wedge\Omega(\Ysp_2),
  \label{c3:gauge-r:Eq}\\
F_{\left(p_s+2\right)}&=&\frac{2}{\sqrt{N_s}}
h_r^{4/N_r}\,d\left[h^{-1}_s(x, v)\right]\wedge\Omega(\Xsp)
\wedge\Omega(\Ysp_1),
  \label{c3:gauge-s:Eq}
}
where $\Omega(\Xsp)$, $\Omega(\Ysp_1)$, and $\Omega(\Ysp_2)$
are the volume $(p+1)$-, $(p_s-p)$-, and $(p_r-p)$-forms, respectively.

If we use the ansatz of metric and fields, 
the field equations give the intersection rule $\chi=-2$
 \cite{Gauntlett:1996pb, Minamitsuji:2010kb}. This is 
different from the usual rule which is obtained in the cases (I) and (II).
From the intersection rule $\chi=-2$, it is easy to show that the field
equations lead
\Eqrsubl{c3:Einstein:Eq}{
&&R_{\mu\nu}(\Xsp)=0,~~~~R_{ij}(\Ysp_1)=0,~~~~
R_{mn}(\Ysp_2)=0,~~~~R_{ab}(\Zsp)=0,
   \label{c3:Ricci:Eq}\\
&&h_r=h_0(x)+h_1(y),~~~~h_s=k_0(x)+k_1(v),
   \label{c3:h:Eq}\\
&&D_{\mu}D_{\nu}h_0=0, ~~~\pd_{\mu}h_0\left[\left(1-\frac{4}{N_r}\right)
\pd_{\nu}\ln h_r-\frac{4}{N_s}\pd_{\nu}\ln h_s\right]=0,
~~~\triangle_{\Ysp_1}h_1=0,
   \label{c3:warp2:Eq}\\
&&D_{\mu}D_{\nu}k_0=0,~~~\pd_{\mu}k_0\left[\left(1-\frac{4}{N_s}\right)
\pd_{\nu}\ln h_s-\frac{4}{N_r}\pd_{\nu}\ln h_r\right]=0,
~~~\triangle_{\Ysp_2}k_1=0\,,
   \label{c3:warp3:Eq}
 }
where $\triangle_{\Ysp_1}$ and $\triangle_{\Ysp_2}$ are the Laplace
operators on the spaces of $\Ysp_1$ and $\Ysp_2$, respectively.
The functions $h_1$ and $k_1$ are nontrivial if
$F_{\left(p_r+2\right)}\ne 0$ and $F_{\left(p_s+2\right)}\ne 0$.
There is no dynamical solution for 
$\frac{1}{N_r}+\frac{1}{N_s}= \frac{1}{4}$
because of $h_r\ne h_s$.

In the case of $\frac{1}{N_r}+\frac{1}{N_s}\neq \frac{1}{4}$,
as in the cases (I) and (II),
if $\partial_{\mu}k_0=0$
there is no solution for $h_0(x)$
such as $\partial_{\mu}h_0\neq 0$ unless $N_r=4$.
Now we focus on a special case with the conditions,
\Eq{
q_{\mu\nu}=\eta_{\mu\nu}\,,~~~\gamma_{ij}=\delta_{ij}\,,~~~
w_{mn}=\delta_{mn}\,,~~~u_{ab}=\delta_{ab}\,,~~~~N_r=4\,,
 \label{c3:flat:Eq}
 }
where $\eta_{\mu\nu}$ is the $(p+1)$-dimensional
Minkowski metric and $\delta_{ij}$, $\delta_{mn}$, $\delta_{ab}$ denote 
the $(p_s-p)$-, $(p_r-p)$-, and $(D+p-p_r-p_s-1)$-dimensional 
Euclidean metrics, respectively.
Upon setting $\pd_{\mu}h_s=0$, 
the solution for $h_r$ and $h_s$ can be written explicitly as
\Eqrsubl{c3:solution:Eq}{
h_r(x, y)&=&A_{\mu}x^{\mu}+B
+\sum_{\ell}\frac{M_\ell}{|\bm y-\bm y_\ell|^{p_s-p-2}},
 \label{c3:solution-r:Eq}\\
h_s(v)&=&C+\sum_{c}\frac{M_c}{|\bm v-\bm v_c|^{p_r-p-2}},
 \label{c3:solution-s:Eq}
}
where $A_{\mu}$, $B$, $C$, $\bm y_\ell$, $\bm v_c$,
$M_\ell$ and $M_c$ denote constant parameters.
Thus, in our time-dependent generalization of intersecting brane solutions,
only D-, NS-branes as well as M-branes can be time dependent
in $D=11$ and $D=10$ supergravities. 
There is also a single brane solution which only exists in the Euclidean
formulation is the (-1)-brane, or D-instanton.
The intersections for the RR-charged D-instantons are thus expressed as 
\Eq{
{\rm D}(-1)\cap {\rm D}p = \frac{1}{2}(p-9).
  \label{c3:int RD:Eq}
}

For $D=6$, we can construct the dynamical intersecting brane solutions 
in the NSS and the Romans theories with a vanishing cosmological constant. 
However, the number of intersections involving the
1-brane and 0-brane are $p=-3$ for the NSS theory
and $p=-2$ for the ${\cal N}=4^{g}$ class of the Romans theory,
respectively. Therefore, both of them are classically meaningless. 

\subsection{Cosmology}
\label{subsec:cosmology}

Let us consider the dynamical solutions for
the $p_r$- and $p_s$-brane system which appears in 
the $D$-dimensional theory. 
In this section, we apply the above solutions to
the four-dimensional cosmology.  
We assume an isotropic and homogeneous three-space 
in the four-dimensional spacetime.
We also assume that the $(p+1)$-dimensional spacetime is 
the Minkowski spacetime with $q_{\mu\nu}(\Xsp)=\eta_{\mu\nu}(\Xsp)$, and
does not depend on the coordinates of $\Xsp$ space except for the time.

The $D$-dimensional metric (\ref{c1:metric:Eq}) can be written as
\Eq{
ds^2=-hdt^2+ds^2(\tilde{\Xsp})+ds^2(\Ysp_1)+
ds^2(\Ysp_2)+ds^2(\Zsp),
   \label{c:metric:Eq}
}
where the metrics $ds^2(\tilde{\Xsp})$, $ds^2(\Ysp_1)$, $ds^2(\Ysp_2)$, 
$ds^2(\Zsp)$ are given by 
\Eqrsubl{c:metric1:Eq}{
ds^2(\tilde{\Xsp})&\equiv&h\delta_{PQ}(\tilde{\Xsp})d\theta^{P}d\theta^{Q},\\
ds^2(\Ysp_1)&\equiv&h^{b_r}_rh_s^{a_s}\gamma_{ij}(\Ysp_1)dy^idy^j,\\
ds^2(\Ysp_2)&\equiv&h^{a_r}_rh_s^{b_s}w_{mn}(\Ysp_2)dv^{m}dv^{n},\\
ds^2(\Zsp)&\equiv&h^{b_r}_rh_s^{b_s}u_{ab}(\Zsp)dz^adz^b,\\
h&\equiv&h^{a_r}_rh_s^{a_s}.
 }
Here, $\theta^P$ is the coordinate of the $p$-dimensional Euclid
space $\tilde{\Xsp}$, and $\delta_{PQ}(\tilde{\Xsp})$ is the 
$p$-dimensional Euclidean metric. 

We focus on the case $h_r= h_s$ and 
$\frac{1}{N_r}+\frac{1}{N_s}=\frac{1}{4}$ and 
set $h_r=At+h_1$. The $D$-dimensional metric
(\ref{c:metric:Eq}) can be expressed as
\Eqr{
ds^2&=&\left[1+\left(\frac{\tau}{\tau_0}\right)^{-2/(a_r+a_s+2)}
h_1\right]^{a_r+a_s}
\left[-d\tau^2+\left(\frac{\tau}{\tau_0}\right)^{2(a_r+a_s)/(a_r+a_s+2)}
\delta_{PQ}(\tilde{\Xsp})d\theta^Pd\theta^Q\right.\nn\\
&&\hspace{-0.7cm}
+\left\{1+\left(\frac{\tau}{\tau_0}\right)^{-2/(a_r+a_s+2)}h_1\right\}
^{4/N_r}
\left(\frac{\tau}{\tau_0}\right)^{2(a_r+a_s+4/N_r)/(a_r+a_s+2)}
\gamma_{ij}(\Ysp_1)dy^idy^j\nn\\
&&\hspace{-0.7cm}
+\left\{1+\left(\frac{\tau}{\tau_0}\right)^{-2/(a_r+a_s+2)}h_1\right\}
^{1-4/N_r}\left(\frac{\tau}{\tau_0}\right)^{2(a_r+a_s+1-4/N_r)/(a_r+a_s+2)}
w_{mn}(\Ysp_2)dv^mdv^n\nn\\
&&\left.\hspace{-0.7cm}
+\left\{1+\left(\frac{\tau}{\tau_0}\right)^{-2/(a_r+a_s+2)}
h_1\right\}
\left(\frac{\tau}{\tau_0}\right)^{2(a_r+a_s+1)/
(a_r+a_s+2)}
u_{ab}(\Zsp)dz^adz^b\right],
   \label{c:metric-a2-2:Eq}
}
where the cosmic time $\tau$ is defined as
\Eq{
\frac{\tau}{\tau_0}=\left(At\right)^{(a_r+a_s+2)/2},~~~~\tau_0=
\frac{2}{\left(a_r+a_s+2\right)A}.
   \label{c:c-time2:Eq}
}
The $D$-dimensional metric \eqref{c:metric-a2-2:Eq} implies that  
the power of the scale factor in the fastest expanding case is 
\Eq{
\frac{a_r+a_s+1}{a_r+a_s+2}=\frac{p_r+1}{D+p_r-1}<1,~~~~~~{\rm for}~~D>2,
}
for $p_r=p_s$ and $N_r=N_s=8$. 
Then, it is impossible to find the cosmological model that our Universe 
exhibits an accelerating expansion.

Now let us consider the lower-dimensional effective theory. 
We compactify $d(\equiv d_1+d_2+d_3+d_4)$ dimensions to fit our Universe,
where $d_1$, $d_2$, $d_3$ and $d_4$ are the compactified dimensions with
respect to the $\tilde{\Xsp}$, $\Ysp_1$, $\Ysp_2$ and $\Zsp$ spaces.
The metric (\ref{c:metric:Eq}) is then given by
\Eq{
ds^2=ds^2(\Msp)+ds^2(\Nsp),
   \label{c:metric2-2:Eq}
}
where $ds^2(\Msp)$ is the $(D-d)$-dimensional metric and
$ds^2(\Nsp)$ denotes the metric of compactified dimensions.

In order to rewrite the $(D-d)$-dimensional metric in the Einstein frame, 
we use the conformal transformation:
\Eq{
ds^2(\Msp)=h_r^Bds^2(\bar{\Msp})\,.
}
where $B$ is defined by 
\Eq{
B=-\frac{(a_r+a_s)d+d_3+d_4+\frac{4(d_2-d_3)}{N_r}}{D-d-2}.
}
Then the $(D-d)$-dimensional metric in the Einstein frame is given by 
\Eqr{
ds^2(\bar{\Msp})&=&h_r^{B'}\left[-d\tau^2+\delta_{P'Q'}
(\tilde{\Xsp}')d\theta^{P'}d\theta^{Q'}
+h_r^{4/N_r}\gamma_{k'l'}({\Ysp_1}')dy^{k'}dy^{l'}\right.\nn\\
&&\left.+h_r^{1-4/N_r}w_{m'n'}({\Ysp_2}')dv^{m'}dv^{n'}
+h_ru_{a'b'}({\Zsp}')dz^{a'}dz^{b'}\right],
  \label{c:metric-m-2:Eq}
}
where $B'$ is defined by $B'=-B+a_r+a_s$,
and $\tilde{\Xsp}'$, ${\Ysp_1}'$, ${\Ysp_2}'$ and ${\Zsp}'$ are 
the $(p-d_1)$-, $(p_s-p-d_2)$-, $(p_r-p-d_3)$-, and
$(D+p-p_r-p_s-1-d_4)$-dimensional spaces, respectively.

Upon setting $h_r=At+h_1$, the metric (\ref{c:metric-m-2:Eq}) is replaced as
\Eqr{
ds^2(\bar{\Msp})&=&\left[1+\left(\frac{\tau}{\tau_0}\right)
^{-2/(B'+2)}h_1\right]^{B'}\left[-d\tau^2+
\left(\frac{\tau}{\tau_0}\right)^{2B'/(B'+2)}
\delta_{P'Q'}(\tilde{\Xsp}')d\theta^{P'}d\theta^{Q'}\right.\nn\\
&&\hspace{-0.5cm}
+\left\{1+\left(\frac{\tau}{\tau_0}\right)^{-2/(B'+2)}h_1\right\}^{4/N_r}
\left(\frac{\tau}{\tau_0}\right)^{2(B'+4/N_r)/(B'+2)}
\gamma_{k'l'}({\Ysp_1}')dy^{k'}dy^{l'}\nn\\
&&\hspace{-0.5cm}
+\left\{1+\left(\frac{\tau}{\tau_0}\right)^{-2/(B'+2)}h_1\right\}^{1-4/N_r}
\left(\frac{\tau}{\tau_0}\right)^{2(B'+1-4/N_r)/(B'+2)}
w_{m'n'}({\Ysp_2}')dv^{m'}dv^{n'}\nn\\
&&\left.\hspace{-0.5cm}
+\left\{1+\left(\frac{\tau}{\tau_0}\right)^{-2/(B'+2)}
h_1\right\}
\left(\frac{\tau}{\tau_0}\right)^{2(B'+1)/(B'+2)}
u_{a'b'}({\Zsp}')dz^{a'}dz^{b'}\right],
  \label{c:metric-mr-2:Eq}
}
where the cosmic time $\tau$ can be expressed as 
\Eq{
\frac{\tau}{\tau_0}=\left(At\right)^{(B'+2)/2},~~~~\tau_0=
\frac{2}{\left(B'+2\right)A}.
}
For the Einstein frame, the power of the scale factor 
in the fastest expanding case is also given by 
\Eq{
\frac{B'+1}{B'+2}<1,~~~~~~{\rm for}~~D>d+2,~~~~d>0.
   \label{c:E-power:Eq}
}
Therefore, we cannot find the solution which
exhibits an accelerating expansion of our Universe. 

The Friedmann-Lemaitre-Robertson-Walker (FLRW) 
cosmological solutions with an isotropic 
and homogeneous three-space for the solutions 
(\ref{c:metric-mr-2:Eq}) are listed in 
Table~\ref{table_2} for the intersection of 0-brane and 
1-brane system in the six-dimensional Romans theory. 
The power exponents of the scale factor of possible four-dimensional 
cosmological models can be expressed as  
$a(\tilde{\Msp})\propto \tau^{\lambda(\tilde{\Msp})}$,
where $\tilde{\Msp}$ denotes the spatial part of the spacetime $\Msp$, 
$\tau$ denotes the cosmic time, and $a(\tilde{\Msp})$ and
$a_{\rm E}(\tilde{\Msp})$ are the scale factors of the space
$\tilde{\Msp}$ in Jordan and Einstein frames with the exponents carrying
the same suffices, respectively.

The obtained expansion law is simple because 
the time dependence in the metric comes from only one
brane in the intersections. Thus one may have to
compactify the vacuum bulk space as well as the brane world volume 
to find an expanding universe. 
However, we find that the fastest expanding case in the Jordan frame
has the power $\lambda(\tilde{\Msp})<1/2$.  
Hence, the solutions cannot give a realistic expansion law like that 
in the matter-dominated era ($a\propto \tau^{2/3}$)
or that in the radiation-dominated era ($a\propto \tau^{1/2}$).

When we derive the four-dimensional effective theory in the Einstein frame
after compactifying the extra dimensions, 
the power exponents are different depending on 
how we compactify the extra dimensions even within one solution.
For the intersection of 0-brane and 1-brane 
in the six-dimensional Romans theory, 
we find the power exponent of the fastest expansion of 
our four-dimensional Universe in the Einstein frame in 
Table \ref{table_7}. Unfortunately, the expansion is too small again.
Therefore, in order to find a realistic expansion
of the Universe in this type of models, one has to include
additional ``matter" fields on the brane.



\section{The intersection of $n$ branes in $D$-dimensional theory}
  \label{sec:ni}

\subsection{Theory}

Let us consider a gravitational theory with the metric $g_{MN}$, 
scalar field $\phi$, and antisymmetric tensor fields of rank $(p_I+2)$, 
where $I$ denotes the type of the corresponding branes.
If the expectation values of fermionic fields are assumed to be zero, 
the most general action for the intersecting-brane system is given by
\Eq{
S=\frac{1}{2\kappa^2}\int \left[R\ast{\bf 1}_D
 -\frac{1}{2}\ast d\phi \wedge d\phi
 -\sum_I \frac{1}{2(p_I+2)!}\e^{\epsilon_Ic_I\phi}\ast F_{(p_I+2)}
 \wedge F_{(p_I+2)}\right],
\label{ni:action:Eq}
}
where $\kappa^2$ is the $D$-dimensional gravitational constant,
$\ast$ denotes the Hodge dual operator in the $D$-dimensional spacetime,
$c_I$, $\epsilon_I$ are constants given by
\Eqr{
c_I^2&=&N_I-\frac{2(p_I+1)(D-p_I-3)}{D-2}, 
\label{ni:c:Eq}\\
\epsilon_I&=&\left\{
\begin{array}{cc}
 + &~{\rm for~the~electric~brane}
\\
 - &~~~{\rm for~the~magnetic~brane}.
\end{array} \right. 
\label{ni:epsilon:Eq}
}

The $D$-dimensional action (\ref{ni:action:Eq}) gives field equations
\Eqrsubl{ni:equations:Eq}{
&&R_{MN}=\frac{1}{2}\pd_M\phi \pd_N \phi
 +\frac{1}{2}\sum_I\frac{1}{(p_I+2)!}\e^{\epsilon_Ic_I\phi}\nn\\
&&~~~~~~~~\times
\left[(p_I+2)F_{MA_2\cdots A_{p_I+2}} {F_N}^{A_2\cdots A_{p_I+2}}
-\frac{p_I+1}{D-2}g_{MN} F^2_{(p_I+2)}\right],
\label{ni:Einstein:Eq}\\
&&d\ast d\phi=\frac{1}{2}\sum_I\frac{\epsilon_Ic_I}{(p_I+2)!}
\e^{\epsilon_Ic_I\phi}\ast F_{(p_I+2)}\wedge F_{(p_I+2)},
 \label{ni:scalar:Eq}\\
&&d\left[\e^{\epsilon_Ic_I\phi}\ast F_{(p_I+2)}\right]=0.
   \label{ni:gauge:Eq}
}

\subsection{Solutions}

We suppose the following ansatz of the $D$-dimensional metric
\Eqr{
\hspace{-10mm}
ds^2=-{\cal A}(t, z)dt^2
+\sum_{\alpha=1}^p{\cal B}^{(\alpha)}(t, z)(dx^{\alpha})^2 
+{\cal C}(t, z)u_{ij}(\Zsp) dz^i dz^j,
 \label{ni:metric:Eq}
}
where
$u_{ij}(\Zsp)$ denotes the metric of the $(D-p-1)$-dimensional $\Zsp$ space 
depending only on the $(D-p-1)$-dimensional coordinates $z^i$.
The functions ${\cal A}$, ${\cal B}^{(\alpha)}$ and ${\cal C}$ are given by
\Eq{
{\cal A}=\prod_I\left[h_I(t,z)\right]^{a_I},~~~
{\cal B}^{(\alpha)}=\prod_I\left[h_I(t,z)\right]^{\delta^{(\alpha)}_I},~~~
{\cal C}=\prod_I\left[h_I(t,z)\right]^{b_I}\,,
}
where $h_I(t,z)$, which depends on $t$ and $z^i$, denotes
a straightforward generalization of the harmonic function associated
with the brane $I$ in a static brane system~\cite{Argurio:1997gt} and 
the parameters $a_I$, $b_I$ and $\delta^{(\alpha)}_I$ are defined by
\Eq{
a_I=-\frac{4(D-p_I-3)}{N_I(D-2)},~~~~b_I=\frac{4(p_I+1)}{N_I(D-2)},~~~~
\delta^{(\alpha)}_I=\left\{
\begin{array}{cc}
a_I&~{\rm for}~~\alpha\in I\\ b_I &~{\rm for}~~\alpha
\in \hspace{-.8em}/ I
\end{array} \right. .
 \label{ni:paremeter:Eq}
}

The scalar field $\phi$ and
the gauge field strength $F_{(p+2)}$ are assumed to be
\Eq{
\e^{\phi}=\prod_Ih_I^{2\epsilon_Ic_I/N_I},~~~
F_{(p_I+2)}=\frac{2}{\sqrt{N_I}}d(h_I^{-1})\wedge\Omega(\Xsp_I),
  \label{ni:fields:Eq}
}
where $\Xsp_I$ is the space associated with the brane $I$, 
and $
\Omega(\Xsp_I)=dt\wedge dx^{p_1}\wedge \cdots \wedge
dx^{p_I}$  is the volume $(p_I+1)$-form.

If we set~\cite{Argurio:1997gt}
\Eq{
{\cal A}^{(D-p-3)}
 \prod_{\alpha=1}^p\,{\cal B}^{(\alpha)} \,{\cal C}=1 \,,~~~~~~~
{\cal A}^{-1}\prod_{\alpha \in I}\left({\cal B}^{(\alpha)}\right)^{-1}
\e^{\epsilon_Ic_I\phi}=h^2_I\,,
   \label{ni:extremal2:Eq}
}
the Einstein equations \Eqref{ni:Einstein:Eq} then give
\Eqrsubl{ni:cEinstein:Eq}{
&&
\sum_{I,I'}\left(\frac{2}{N_I}\delta_{II'}-M_{II'}
\right)\pd_t\ln h_I\pd_t\ln h_{I'}\nn\\
&&~~~~+\frac{1}{2}\sum_{I}b_I\left[\left(1-\frac{4}{N_I}\right)\pd_t\ln h_I
-\sum_{I'\ne I}\frac{4}{N_{I'}}\pd_t\ln h_{I'}\right]\pd_t\ln h_I
\nn\\
&&~~~~-\frac{1}{2}\sum_{I}\left(\frac{4}{N_I}+b_I\right)h_I^{-1}\pd_t^2h_I
+\frac{1}{2}\prod_{I'}h_{I'}^{-4/N_{I'}}\sum_{I}a_{I}
h_{I}^{-1}\lap_{\Zsp}h_{I}=0,
   \label{ni:Einstein tt:Eq}\\
&&\sum_I\frac{2}{N_I}h^{-1}_I\pd_t\pd_ih_I+\sum_{I,I'}
\left(M_{II'}-\frac{2}{N_{I}}\delta_{II'}
\right)\pd_t\ln h_I\pd_i\ln h_{I'}=0,
   \label{ni:Einstein ti:Eq}\\
&&\prod_{J'}h_{J'}^{-a_{J'}}
\sum_{\gamma}\prod_{J}h_{J}^{\delta_J^{(\gamma)}}
\sum_{I} \delta^{(\gamma)}_I\left[h_I^{-1}\pd_t^2h_I
-\left\{\left(1-\frac{4}{N_I}\right)\pd_t\ln h_I
\right.\right.\nn\\
&&\left.\left.~~~-\sum_{I'\ne I}\frac{4}{N_{I'}}
\pd_t\ln h_{I'}\right\}\pd_t\ln h_I\right]
-\prod_{J'}h_{J'}^{-b_{J'}}
\sum_{\gamma}\prod_{J}h_{J}^{\delta_J^{(\gamma)}}\sum_I
\delta^{(\gamma)}_Ih_I^{-1}\lap_{\Zsp}h_I=0,
   \label{ni:Einstein ab:Eq}\\
&&R_{ij}(\Zsp)+\frac{1}{2}u_{ij}\prod_Jh_J^{4/N_J}\sum_{I}
b_I\left[ h_I^{-1}\pd_t^2h_I
-\left\{\left(1-\frac{4}{N_I}\right)\pd_t\ln h_I
-\sum_{I'\ne I}\frac{4}{N_{I'}}\pd_t\ln h_{I'}\right\}\pd_t\ln h_I\right]\nn\\
&&~~~~-\frac{1}{2}u_{ij}\sum_Ib_I h_I^{-1}\lap_{\Zsp}h_I
-\sum_{I,I'}\left(M_{II'}-\frac{2}{N_I}\delta_{II'}
\right)\pd_i\ln h_I\pd_j\ln h_{I'}=0 ,
    \label{ni:Einstein ij:Eq}
}
where $R_{ij}(\Zsp)$ is the Ricci tensor constructed from the metric $u_{ij}$,
and $M_{II'}$ is defined by
\Eqr{
&&M_{II'}\equiv \frac{1}{4}\left[a_Ia_{I'}
+\sum_{\alpha}\delta^{(\alpha)}_I\delta^{(\alpha)}_{I'}
+(D-p-3)b_Ib_{I'}\right] 
+\frac{2}{N_IN_{I'}}\epsilon_I\epsilon_{I'}c_Ic_{I'}
\,.
\label{ni:Ein1:Eq}
}

For \Eqref{ni:Einstein ti:Eq}, we can rewrite this as
\Eq{
\sum_{I,I'}\left[ M_{II'}+\frac{2}{N_{I}}
\delta_{II'}\frac{\pd_t\pd_i \ln h_I}{\pd_t \ln h_I \pd_i \ln h_I} \right]
\pd_t \ln h_I \pd_i \ln h_{I'}=0.
\label{ni:Ein2:Eq}
}
The second term in the square bracket of (\ref{ni:Ein2:Eq}) must be 
constant to satisfy this equation for arbitrary coordinate values and 
independent functions $h_I$. Then we find 
\Eq{
\frac{\pd_t\pd_i \ln h_I}{\pd_t \ln h_I \pd_i \ln h_I}=k_I
\,.
\label{ni:Ein3:Eq}
}
Hence, \Eqref{ni:Ein2:Eq} leads to 
\Eq{
M_{II'}+\frac{2}{N_I}k_I \delta_{II'}=0.
\label{ni:const1:Eq}
}
In terms of Eqs.~(\ref{ni:c:Eq}), (\ref{ni:paremeter:Eq}) and 
(\ref{ni:Ein1:Eq}), we find
\Eqr{
M_{II} &=&\frac{1}{4}
 \left[(p_I+1)a_I^2 + (p-p_I)b_I^2
+(D-p-3)b_I^2\right] +\frac{2}{N_I^2}c_I^2 \nonumber \\
&=& \frac{2}{N_I}.
\label{ni:m:Eq}
}
From the Eq.~(\ref{ni:m:Eq}), the constant $k_I$ in~\Eqref{ni:const1:Eq} 
obeys $k_I=-1$, namely
\Eq{
M_{II'}= \frac{2}{N_{I'}}\delta_{II'}.
\label{ni:m2:Eq}
}
Then \Eqref{ni:Ein3:Eq} gives
\Eq{
\pd_t \pd_i[h_I(t,z)]=0
\,.
}
As a result, the function $h_I$ have to be separable as
\Eq{
h_I(t, z)= K_I(t)+H_I(z)
\,.
  \label{ni:hi:Eq}
}

If we set $I\neq I'$, ~\Eqref{ni:m2:Eq} gives the intersection rule
on the dimension $\bar{p}$ of the intersection for each pair of branes
$I$ and $I'$ $(\bar{p}\leq p_I, p_{I'})$~\cite{Gauntlett:1996pb,
Tseytlin:1996hi, Argurio:1997gt,Ohta:1997gw}:
\Eq{
\bar{p}=\frac{(p_I+1)(p_{I'}+1)}{D-2}-1-
\frac{1}{2}\epsilon_Ic_I\epsilon_{I'}c_{I'}.
\label{ni:isr:Eq}
}

Let us next consider the gauge field.
Using the ansatz~\Eqref{ni:fields:Eq} for electric background,
we find
\Eq{
h_I^{-1}(2\pd_i \ln h_I \pd_j \ln h_I
+ h_I^{-1}\pd_i\pd_j h_I)
dz^i\wedge dz^j\wedge\Omega(\Xsp_I)=0.
}
Then, the Bianchi identity is automatically satisfied.
The equation of motion for the gauge field also becomes
\Eqr{
d\left[\pd_iH_I \left(\ast_{\Zsp}dz^i\right)\wedge
\ast_{\Xsp}\Omega(\Xsp_I)\right]=0,
 }
where $\ast_{\Xsp}$, $\ast_{\Zsp}$ are the Hodge dual
operator on $\Xsp(\equiv \cup_{I}X_I)$ and $\Zsp$, respectively,
and we have used Eq.~(\ref{ni:extremal2:Eq}). 
Thus, we again obtain the condition~\Eqref{ni:hi:Eq} and
\Eq{
\lap_{\Zsp}H_I=0\,,
\label{ni:gauge2:Eq}
}
where we used Eq.~(\ref{ni:hi:Eq}). 
Although the roles of the Bianchi identity and field equations are
interchanged for magnetic ansatz~\cite{Argurio:1997gt,Ohta:1997gw},
the net result is the same.

Finally we consider the scalar field equation.
Substituting the scalar field and the gauge field in 
\Eqref{ni:fields:Eq}, and the function \Eqref{ni:hi:Eq}
into the scalar field equation 
\Eqref{ni:scalar:Eq}, we obtain
\Eqr{
&&-\prod_{I''}h_{I''}^{-a_{I''}}
\sum_{I}\frac{1}{N_I}\epsilon_Ic_I\left[h_I^{-1}\pd_t^2K_I
-\left\{\left(1-\frac{4}{N_I}\right)\pd_t\ln h_I-
\sum_{I'\ne I}\frac{4}{N_{I'}}\pd_t\ln h_{I'}\right\}\pd_t\ln h_I
\right]\nn\\
&&~~~~~
+\prod_{I''}h_{I''}^{-b_{I''}}\sum_I\frac{1}{N_I}h_I^{-1}
\epsilon_Ic_I\lap_{\Zsp} H_I=0.
  \label{ni:scalar2:Eq}
}
The equation (\ref{ni:scalar2:Eq}) is satisfied if
\Eqrsubl{ni:enough:Eq}{
&&\pd_t^2K_I=0,
\label{ni:eq_KI:Eq}
\\
&&\triangle_{\Zsp}H_I=0,
   \label{5}\\
&&\sum_{I}\frac{1}{N_I}
\epsilon_Ic_I\left[\left(1-\frac{4}{N_I}\right)\pd_t\ln h_I-
\sum_{I'\ne I}\frac{4}{N_{I'}}\pd_t\ln h_{I'}
\right]=0.
\label{ni:scalar3:Eq}
}
Equation (\ref{ni:eq_KI:Eq}) gives
$K_I=A_I t+B_I$, where $A_I$ and $B_I$ are integration constants.

{\bf (A):}
Let us first consider the case that we take all functions to be equal:
\Eq{
h_I(t,z)=h(t,z)\equiv K(t)+H(z),~~~~N_I=N_{I'}=N.
    \label{ni:h:Eq}
}
We can find the solutions if the function $h$ and $N$ satisfy
\Eq{
K(t)=A\, t+B,~~~~N=4\ell,
\label{ni:sol_K:Eq}
}
where $\ell$ denotes the number of the functions $h_I$.
Then the remaining
Einstein equations~\Eqref{ni:cEinstein:Eq} are 
\Eq{
R_{ij}(\Zsp)=0.
   \label{ni:Ricci:Eq}
}
Now we assume 
\Eq{
\quad u_{ij}=\delta_{ij}\,,
 \label{ni:flat:Eq}
}
where $\delta_{ij}$ is the $(D-p-1)$-dimensional Euclidean metric.
In this case, the solution for $h$ can be obtained explicitly as
\Eq{
h(t, z)=At+B +\sum_{k}\frac{Q_{k}}{|\bm{z}- \bm{z}_{k}|^{D-p-3}},
}
where  $Q_{k}$'s are constant parameters and $\bm{z}_{k}$
represents the positions of the branes in Z space, 
$\bm z_K$ is constant vector representing the positions of the branes. 
Since the functions coincide, the locations of the 
brane will also coincide. 
This physically means that all branes have the same 
total amount of charge at same position. 

Let us consider the intersection rule in the $D$-dimensional
theory. If we choose $p_I=\tilde{p}$ for all $p_I$, 
the intersection rule Eq.~(\ref{ni:isr:Eq}) leads to
\Eq{
\bar{p}=\tilde{p}-2\ell.
    \label{ni:chi2:Eq}
}
Then, we get the intersection involving two $\tilde{p}$-brane
\Eq{
\tilde{p}\cap \tilde{p}=\tilde{p}-2\ell.
   \label{ni:int:Eq}
}
Equation~\eqref{ni:int:Eq} tells us that the number of
intersection for $\tilde{p}<2\ell$ is negative,
which means that there is no intersecting solution of these brane systems.

If we set $K=0 ~(A=B=0$), 
the metric describes the known static and extremal multi-black hole
solution with black hole charges
$Q_{k}$~\cite{Argurio:1997gt,Ohta:1997gw,Argurio:1998cp}.

{\bf (B):}
Next, we consider the case 
that there is only one function $h_I$ depending on both 
$z^i$ and $t$, which we denote with the 
subscript $\tilde{I}$, and other functions are either dependent on
$z^i$ or constant. We also assume $N_{\tilde{I}}=4$. Then, we have 
\Eqrsubl{ni:N:Eq}{
K_{\tilde{I}}(t)&=&A\, t+B_{\tilde{I}},~~~~N_{\tilde{I}}=4,\\
K_I&=&B_I,~~~~(I\ne I_{\tilde{I}}).
}
If we assume $u_{ij}=\delta_{ij}$, the solution for $H_I$
can be obtained explicitly as 
\Eq{
H_I(z)=1+\sum_{k}\frac{Q_{I, k}}{|\bm{z}- \bm{z}_{k}|^{D-p-3}},
     \label{ni:Hi2:Eq}
}
where  $Q_{I, k}$'s are constant parameters and $\bm{z}_{k}$
represents the positions of the branes in Z space.
We can find the solution (\ref{ni:N:Eq}) for any $N_I$. 
If we choose $N_I=4$, the solutions have already discussed in 
\cite{Maeda:2009zi}.

\subsection{Cosmology}

Let us consider the dynamical solutions for 
the $p_I$-brane system which appears in the $D$-dimensional theory. 
In this section, we apply the above solutions to
the four-dimensional cosmology.  
We assume that the four-dimensional spacetime is 
an isotropic and homogeneous three-space, 
and either the world volume space or the transverse space
can be (a part of) our four-dimensional Universe.

In this section, we focus on the $(D-p-1)$-dimensional 
Euclidean space with $u_{ij}(\Zsp)=\delta_{ij}(\Zsp)$, and 
consider the case that all functions $h_I$ are equal to $h$ 
and all parameter of $N_I$ has the same value $N_I=N=4\ell$, where 
$\ell$ is the number of $p_I$-brane. 
In this case,  the $D$-dimensional metric (\ref{ni:metric:Eq}) 
can be expressed as
\Eq{
ds^2=-h^adt^2+\sum_{\alpha}h^{\delta^{(\alpha)}}
\left(dx^{\alpha}\right)^2+h^b\delta_{ij}(\Zsp)dz^idz^j,
   \label{ni:cmetric:Eq}
}
where $h$ is defined by \eqref{ni:h:Eq} and 
the parameters $a$, $b$, $\delta^{(\alpha)}$ are given by 
\Eq{
a=-\sum_I\frac{D-p_I-3}{\ell(D-2)},~~~~
b=\sum_I\frac{p_I+1}{\ell(D-2)},~~~~
\delta^{(\alpha)}=\sum_I\delta_I^{(\alpha)}.
}
For $K=At$, the metric (\ref{ni:cmetric:Eq}) is thus rewritten as
\Eqr{
ds^2&=&-\left[1+\left(\frac{\tau}{\tau_0}\right)^{-\frac{2}{a+2}}
H\right]^ad\tau^2
+\sum_{\alpha}\left[1+\left(\frac{\tau}{\tau_0}\right)
^{-\frac{2}{a+2}}H\right]^{\delta^{(\alpha)}}
\left(\frac{\tau}{\tau_0}\right)^{\frac{2\delta^{(\alpha)}}{a+2}}
\left(dx^{\alpha}\right)^2\nn\\
&&+\left[1+\left(\frac{\tau}{\tau_0}\right)^{-\frac{2}{a+2}}
H\right]^b\left(\frac{\tau}{\tau_0}\right)^{\frac{2b}{a+2}}
\delta_{ij}(\Zsp)dz^idz^j,
\label{ni:hinden:Eq}
}
where we have introduced the cosmic time $\tau$ defined by
\Eq{
\frac{\tau}{\tau_0}=\left(At\right)^{(a+2)/2},~~~~\tau_0=
\frac{2}{\left(a+2\right)A}.
}
Here, the $H$ is defined by 
\Eq{
H(z)=1+\sum_{k}\frac{Q_{k}}{|\bm{z}- \bm{z}_{k}|^{D-p-3}},
}
where  $Q_{k}$'s are constant parameters and $\bm{z}_{k}$
represents the positions of the branes in Z space.

The $D$-dimensional metric \eqref{ni:hinden:Eq}
implies that  
the power of the scale factor in the fastest expanding case is 
\Eq{
\frac{b}{a+2}=\frac{b}{b+1}<1,
~~~~~~{\rm for}~~D>2.
}
Then, it is impossible to find the cosmological model that our Universe 
exhibits an accelerating expansion.

We compactify $d(\equiv \sum_{\alpha}d_{\alpha}+d_z)$ 
dimensions to fit our Universe, 
where $d_{\alpha}$ and $d_z$ denotes the compactified dimensions with
respect to the relative transverse space, $\Zsp$ spaces.
The metric (\ref{ni:metric:Eq}) is then expressed as
\Eq{
ds^2=ds^2(\Msp)+ds^2(\Nsp),
   \label{ni:metric2:Eq}
}
where $ds^2(\Msp)$ is the metric of $(D-d)$-dimensional spacetime and
$ds^2(\Nsp)$ is the metric of compactified dimensions.

In terms of the conformal transformation
\Eq{
ds^2(\Msp)=h^Bds^2(\bar{\Msp}),
}
the $(D-d)$-dimensional metric can be written in the Einstein frame, 
where $B$ is defined by
\Eq{
B=-\frac{\sum_{\alpha}d_{\alpha}\delta^{(\alpha)}+d_zb}{D-d-2}.
}
Hence, the $(D-d)$-dimensional metric in the Einstein frame is
\Eqr{
ds^2(\bar{\Msp})&=&h^{-B}\left[
-h^adt^2+\sum_{{\alpha}'}h^{\delta^{({\alpha}')}}
\left(dx^{{\alpha}'}\right)^2+h^b\delta_{i'j'}({\Zsp}')dz^{i'}dz^{j'}\right],
  \label{ni:metric-m:Eq}
}
where $x^{{\alpha}'}$ is the coordinate of $(p-d_{\alpha})$-dimensional 
relative transverse space, and ${\Zsp}'$ denote 
$(D-p-1-d_z)$-dimensional spaces, respectively.

If we set $K=At$, the metric (\ref{ni:metric-m:Eq}) is thus rewritten as
\Eqr{
ds^2(\bar{\Msp})&=&
-\left[1+\left(\frac{\tau}{\tau_0}\right)^{-\frac{2}{B'+2}}
H\right]^{B'}d\tau^2\nn\\
&&+\sum_{\alpha'}\left[1+\left(\frac{\tau}{\tau_0}\right)
^{-\frac{2}{B'+2}}H\right]^{-B+\delta^{(\alpha')}}
\left(\frac{\tau}{\tau_0}\right)^
{\frac{2\left(-B+\delta^{(\alpha')}\right)}{B'+2}}
\left(dx^{\alpha'}\right)^2\nn\\
&&+\left[1+\left(\frac{\tau}{\tau_0}\right)^{-\frac{2}{B'+2}}
H\right]^{B'+1}
\left(\frac{\tau}{\tau_0}\right)^{\frac{2\left(B'+1\right)}{B'+2}}
\delta_{i'j'}({\Zsp}')dz^{i'}dz^{j'},
  \label{ni:metric-mr:Eq}
}
where $B'$ is defined by $B'=-B+a$, and
the cosmic time $\tau$ is defined by
\Eq{
\frac{\tau}{\tau_0}=\left(At\right)^{(B'+2)/2},~~~~\tau_0=
\frac{2}{\left(B'+2\right)A}.
}
For the Einstein frame, the power of the scale factor 
in the fastest expanding case is also given by 
\Eq{
\frac{B'+1}{B'+2}<1,~~~~~~{\rm for}~~D>d+2,~~~~d>0.
   \label{ni:E-power:Eq}
}
Then, we cannot find the solution which
exhibits an accelerating expansion 
of our Universe. 

Next we consider the case 
that there is only one function $h_I$ depending on both 
$z^i$ and $t$, which we denote with the 
subscript $\tilde{I}$, and other functions are either dependent on
$z^i$ or constant. We also assume $N_{\tilde{I}}=4$. Then we have 
\Eqr{
\label{ni:metric_N:Eq}
ds^2&=&-\prod_{I\ne \tilde{I}}h_I^{a_I}
\left[1+\left(\frac{\tau}{\tau_0}\right)^{-\frac{2}{a_{\tilde{I}}+2}}
H_{\tilde{I}}\right]^{a_{\tilde{I}}}d\tau^2\nn\\
&&
+\sum_{\alpha}\prod_{I\ne \tilde{I}}h_I^{\delta_I^{(\alpha)}}
\left[1+\left(\frac{\tau}{\tau_0}\right)
^{-\frac{2}{a_{\tilde{I}}+2}}H_{\tilde{I}}\right]^
{\delta_{\tilde{I}}^{(\alpha)}}
\left(\frac{\tau}{\tau_0}\right)^{\frac{2\delta_{\tilde{I}}^{(\alpha)}}
{a_{\tilde I}+2}}
\left(dx^{\alpha}\right)^2\nn\\
&&+\prod_{I\ne \tilde{I}}h_I^{b_I}
\left[1+\left(\frac{\tau}{\tau_0}\right)^{-\frac{2}{a_{\tilde{I}}+2}}
H_{\tilde{I}}\right]^{b_{\tilde{I}}}
\left(\frac{\tau}{\tau_0}\right)^{\frac{2b_{\tilde{I}}}{a_{\tilde{I}}+2}}
\delta_{ij}(\Zsp)dz^idz^j,
}
where the function $H_{\tilde{I}}$ is defined by \eqref{ni:Hi2:Eq}, and 
we have introduced the cosmic time $\tau$ defined as
\Eq{
\frac{\tau}{\tau_0}=\left(At\right)^{(a_{\tilde{I}}+2)/2},~~~~\tau_0=
\frac{2}{\left(a_{\tilde{I}}+2\right)A}.
}

The $D$-dimensional metric (\ref{ni:metric_N:Eq}) implies that  
the power of the scale factor in the fastest expanding case is 
\Eq{
\frac{b_{\tilde{I}}}{a_{\tilde{I}}+2}=
\frac{p_{\tilde{I}}+1}{D+p_{\tilde{I}}-1}<1,
~~~~~~{\rm for}~~D>2.
}
Then, it is impossible to find the cosmological model that our Universe 
exhibits an accelerating expansion.

We compactify $d(\equiv \sum_{\alpha}d_{\alpha}+d_z)$ 
dimensions to fit our Universe, 
where $d_{\alpha}$ and $d_z$ denotes the compactified dimensions with
respect to the relative transverse space, $\Zsp$ spaces.
The metric (\ref{ni:metric:Eq}) is then written by \eqref{ni:metric2:Eq}. 
By the conformal transformation
\Eq{
ds^2(\Msp)=h_{\tilde{I}}^{B_{\tilde{I}}}\prod_{I\ne \tilde{I}}
h_I^{C_I}ds^2(\bar{\Msp}),
}
the $(D-d)$-dimensional metric can be written in the Einstein frame, where 
$B_{\tilde{I}}$ and $C_I$ are given by
\Eq{
B_{\tilde{I}}=-\frac{\sum_{\alpha}d_{\alpha}
{\delta_{\tilde{I}}}^{(\alpha)}+d_zb_{\tilde{I}}}{D-d-2},~~~~
C_I=-\frac{\sum_{\alpha}d_{\alpha}\delta_I^{(\alpha)}+d_zb_I}{D-d-2}.
}
Then, the $(D-d)$-dimensional metric in the Einstein frame is rewritten by
\Eqr{
ds^2(\bar{\Msp})&=&h_{\tilde{I}}^{-B_{\tilde{I}}}\prod_{J\ne \tilde{I}}
{h_J}^{-C_J}\left[
-h^{a_{\tilde{I}}}\prod_{I\ne \tilde{I}}h_I^{a_I}
dt^2+\sum_{{\alpha}'}h_{\tilde{I}}^{\delta_{\tilde{I}}^{({\alpha}')}}
\prod_{I\ne \tilde{I}}h_I^{\delta_I^{({\alpha}')}}
\left(dx^{{\alpha}'}\right)^2\right.\nn\\
&&\left.+h_{\tilde{I}}^{b_{\tilde{I}}}\prod_{I\ne \tilde{I}}h_I^{b_I}
\delta_{i'j'}({\Zsp}')dz^{i'}dz^{j'}\right],
  \label{ni:metric-m2:Eq}
}
where $x^{{\alpha}'}$ is the coordinate of $(p-d_{\alpha})$-dimensional 
relative transverse space, and ${\Zsp}'$ denote 
$(D-p-1-d_z)$-dimensional spaces, respectively.
For $K_{\tilde{I}}=At$, 
the $(D-d)$-dimensional metric in the Einstein frame is 
thus written as
\Eqr{
ds^2(\bar{\Msp})&=&\prod_{I\ne \tilde{I}}h_I^{-C_I}
\left[-\prod_{I\ne \tilde{I}}h_I^{a_I}
\left\{1+\left(\frac{\tau}{\tau_0}\right)^{-\frac{2}{{B'}_{\tilde{I}}+2}}
H_{\tilde{I}}\right\}^{{B'}_{\tilde{I}}}d\tau^2\right.\nn\\
&&+\sum_{\alpha'}\prod_{I\ne \tilde{I}}h_I^{\delta^{(\alpha')}_I}
\left\{1+\left(\frac{\tau}{\tau_0}\right)
^{-\frac{2}{{B'}_{\tilde{I}}+2}}H_{\tilde{I}}\right\}^
{-B_{\tilde{I}}+\delta_{\tilde{I}}^{(\alpha')}}
\left(\frac{\tau}{\tau_0}\right)^{\frac{2\left(-B_{\tilde{I}}
+\delta_{\tilde{I}}^{(\alpha')}\right)}
{{B'}_{\tilde{I}}+2}}
\left(dx^{\alpha'}\right)^2\nn\\
&&\left.+\prod_{I\ne \tilde{I}}h_I^{b_I}
\left\{1+\left(\frac{\tau}{\tau_0}\right)^{-\frac{2}{{B'}_{\tilde{I}}+2}}
H_{\tilde{I}}\right\}^{{B'}_{\tilde{I}}+1}
\left(\frac{\tau}{\tau_0}\right)^{\frac{2\left({B'}_{\tilde{I}}+1\right)}
{{B'}_{\tilde{I}}+2}}
\delta_{i'j'}({\Zsp}')dz^{i'}dz^{j'}\right],
  \label{ni:metric-mr2:Eq}
}
where ${B'}_{\tilde{I}}$ is defined by 
${B'}_{\tilde{I}}=-B_{\tilde{I}}+a_{\tilde{I}}$, and 
the cosmic time $\tau$ is defined by
\Eq{
\frac{\tau}{\tau_0}=\left(At\right)^{({B'}_{\tilde{I}}+2)/2},~~~~\tau_0=
\frac{2}{\left({B'}_{\tilde{I}}+2\right)A}\,.
}
For the Einstein frame, the power of the scale factor 
in the fastest expanding case is also given by 
\Eq{
\frac{{B'}_{\tilde{I}}+1}{{B'}_{\tilde{I}}+2}<1,
~~~~~~{\rm for}~~D>d+2,~~~~d>0.
   \label{ni:E-power2:Eq}
}
Hence, we cannot find the solution which
exhibits an accelerating expansion 
of our Universe.


\section{Discussions}
  \label{sec:discussions}

In the first part of the paper, 
we have seen that dynamical solutions of $p$-brane 
have several remarkable properties. 
If the scalar and gauge fields are related to the functions $h_I$ like 
(\ref{s:fields:Eq}),  
then by counting solutions of the Einstein equations, 
one would construct only the cosmological model of 
decelerating expansion of our Universe. 
We recall that the cosmological constant leads to the accelerating 
expansion which was described somewhat abstractly in Sec.~\ref{sec:cc}.

It appears that the exact forms of the field strengths are 
given by the ansatz (\ref{s:fp:Eq}), 
which depends on the dilaton coupling parameter $N$. The $N=4$ 
case is apparently related to the classical solutions of string theory.
We observed that the dynamical solutions with $N\ne 4$ certainly 
have many attractive properties. 
Firstly, these solutions were obtained by replacing the time-independent 
warp factor of the static solution  
with the time-dependent function. 
The warp factor 
for $N\ne 4$ is the same form as that for $N=4$. 
Secondly, we could not obtain any analytic solution 
of a single $p$-brane with time dependence of the warp factor,
if there is no cosmological constant  
because of the ansatz of the gauge field. 
Since the field strength 
has the component along the time coordinate, the time derivative 
of the warp factor is not permissible in the field strength.
Hence, in the Einstein equations,
the term of the time derivative of the warp factor 
arises only from the Ricci tensor,
and cannot be compensated by 
the scalar and gauge fields, except for $N=4$.

In the case of $N\ne 4$
with a flat transverse space to the brane 
and a positive cosmological constant $\Lambda>0$, 
the Einstein equations give an asymptotically de Sitter solution for a single
2-form field strength.
 To find the solutions to the Einstein equations 
in this way, we need a $D$-dimensional theory with vanishing dilaton 
in which the cosmological constant is related to a field strength. 
This is a generalization of Kastor-Traschen solution in 
four-dimensional Einstein-Maxwell theory. 
We have simply started with $D$-dimensional gravitational theory 
and introduced the cosmological constant with a scalar field that 
preserves time dependence. For the 0-brane in the NSS model of $D=6$, 
an asymptotically Milne solution is obtained. However,
it cannot provide an accelerating universe. 
We have also 
applied the asymptotically de Sitter solution of 
five dimensions
to construct the brane world model. 
We have employed the standard copy and paste 
method to construct a cosmological 3-brane world,
supported by either the tension or induced gravity,
 and embedded into a five-dimensional bulk. 
We have derived the effective gravitational equations
via the junction condition, and shown that the solution gives 
an accelerating expansion on the 3-brane. 
However, in our model there is no natural way
to explain why the bulk cosmological constant is so small. 

In the second part of the paper,
we have discussed the time-dependent intersecting brane solutions. 
For $N_I=4$, which are the parameters in the coupling of the 
field strengths to the 
dilaton, there is only one function $h_I$ depending on both the time 
and coordinates of transverse space. 
All the field strengths in the $D=11$ and $D=10$ supergravities
have $N_I=4$ couplings. 

If all the branes have equal number of world volume dimensions 
and the same charge, it is possible to
get a solution in which all functions $h_I$ depend on both the time 
and the coordinates of overall transverse space.
This turns out to be the only situation where the parameters $N_I$ 
have proper values within the framework of the 
intersecting $p$-brane systems. 

If at least one of branes has $N_{\tilde{I}}=4$, we can construct the
time-dependent solutions even if all other $N_I\ne 4$. 
In this case, only one time-dependent $h_I$ is obtained from 
the brane of $N_{\tilde{I}}=4$. For instance, in the case of 
$D=6$ without a cosmological constant, we have obtained a 
dynamical solution involving 0- and 1-brane in a class of the Romans theory. 
A dynamical intersecting brane system in this class of the Romans theory 
was allowed only for the 1-brane. 

Supposing that our four-dimensional spacetime is located at 
a particular place of the extra spatial dimensions, we have 
obtained expanding FLRW universes. 
The power of the scale factor in these solutions,
however, is too small to give a realistic expansion law even 
in the case that all functions $h_I$ depend on both the time and 
coordinates of overall transverse space. This means that we have to consider 
additional matter on the brane in order to get a realistic expanding 
universe.

As we have observed, there is a serious difficulty in obtaining an 
accelerating expansion from the dynamical intersecting solutions. 
We have discussed the possible solutions 
of field equations for a given scalar and gauge fields 
in Secs.~\ref{sec:two} and \ref{sec:ni}.
For a given choice of ansatz of fields in the $D$-dimensional 
spacetime for the dilaton coupling parameter $c_I$, the functions 
$h_I$ in the metric have a condition corresponding to the 
relation between the warp factors
associated to the parameter $N_I$ in the coupling constant $c_I$. 
In terms of the field equations, 
the functions $h_I$ have a structure of the linear 
combination of the functions $h_0(t)$ and $h_1(z)$. 
The condition for the form of $h_I$ 
to be harmonic function to the transverse space
is not relevant to the choice of $N_I$. 
Though this result is really natural 
in the viewpoint of the extension of 
the static solution, it prevents us from obtaining 
an accelerating expansion because the field equations 
lead to the function $h_0(t)$ depending on the linear function of time. 

Of course, whether this makes sense depends on 
the ansatz of fields associated with $D$-dimensional symmetry. 
A more precise statement with respect to an accelerating expansion  
in the $p$-brane system will be presented in the near future.

\section*{Acknowledgments}
M.M. is grateful for fruitful discussions
during the JGRG 20 and the COSMO/COSPA 2010 held in Japan. 
K.U. would like to thank H. Kodama, M. Sasaki, N. Ohta and T. Okamura 
for continuing encouragement. K.U. is supported 
by Grant-in-Aid for Young Scientists (B) of JSPS Research, under
Contract No. 20740147.




\begin{table}[p]
\caption{\baselineskip 14pt
Intersections of 0-branes and 1-brane of the six-dimensional Romans theory 
with $N_r=4$ for 1-brane and $N_s=2$ for 0-brane in the case (I) and 
(II) are shown. 
Time dependence appears only in 1-brane.
}
\label{twoRomans}
{\scriptsize
\begin{center}
\begin{tabular}{|c||c|c|c|c|c|c|c||c|c|c|
}
\hline
Branes&&0&1&2&3&4&5& $\tilde{\Msp}$ & $\lambda(\tilde{\Msp})$
& $\lambda_{\rm E}(\tilde{\Msp})$
\\
\hline
&1-brane & $\circ$ & $\circ$ & &&& 
& $\Zsp$ & $\lambda(\Zsp)=1/3$ & 
$\lambda_{\rm E}(\Zsp)=
\frac{2-d_3}{6-2d_3-d_4}$
\\
\cline{3-8}
0-brane and 1-brane & 0-brane & $\circ$ &&&&&
&  &  & 
\\
\cline{3-8}
&$x^N$ & $t$ & $v$ & $z^1$ & $z^2$ & $z^3$ & $z^4$ &
 &   & 
\\
\hline
\end{tabular}
\end{center}
}
\label{table_2}
\end{table}

\begin{table}[p]
\caption{\baselineskip 14pt
The power exponent of the fastest expansion in the Einstein frame
for the intersection of 0-brane and 1-brane on 
the six-dimensional Romans theory is shown.
 ``TD" in the table represents which brane is time dependent.
}
\begin{center}
{\scriptsize
\begin{tabular}{|c|c|c||c|c||c||c|}
\hline
Branes& TD &dim$(\Msp)$  &$\tilde{\Msp}$
&$(d_1, d_2, d_3, d_4)$&
$\lambda_{\rm E}(\tilde{\Msp})$ & Case
\\
\hline
0-brane and 1-brane & 1-brane & 5&  Z & (0, 0, 1, 0) &1/4& I \& II
\\
\hline
\end{tabular}
}
\label{table_7}
\end{center}
\end{table}

\end{document}